\documentclass{llncs}

\usepackage[nounderscore]{syntax}
\usepackage{amsmath,amssymb,graphicx,bbm,calc,capt-of,ifthen,theorem}
\usepackage[lined,linesnumbered,boxed]{algorithm2e}
\usepackage{listings}
\usepackage{float}

\newcommand{\texthfo}[1]{Hierarchical Floorplans of Order $#1$ }
\newcommand{\textsgto}[1]{Skewed Generating Trees of Order $#1$ }

\newcommand{\tmem}[1]{{\em #1\/}}
\newcommand{\tmop}[1]{\ensuremath{\operatorname{#1}}}

\newcommand{\tmtextit}[1]{{\itshape{#1}}}
\newtheorem{observation}{Observation}
\newcommand{\tmfloatcontents}{}
\newlength{\tmfloatwidth}
\newcommand{\tmfloat}[5]{
  \renewcommand{\tmfloatcontents}{#4}
  \setlength{\tmfloatwidth}{\widthof{\tmfloatcontents}+1in}
  \ifthenelse{\equal{#2}{small}}
    {\ifthenelse{\lengthtest{\tmfloatwidth > \linewidth}}
      {\setlength{\tmfloatwidth}{\linewidth}}{}}
    {\setlength{\tmfloatwidth}{\linewidth}}
  \begin{minipage}[#1]{\tmfloatwidth}
    \begin{center}
      \tmfloatcontents
      \captionof{#3}{#5}
    \end{center}
  \end{minipage}}

\title{Subclasses of Baxter Permutations Based on Pattern Avoidance}

\author{Shankar Balachandran \and Sajin Koroth}

\institute{\email{shankar,sajin@cse.iitm.ac.in}\\ Department of Computer Science and Engineering,\\ Indian Institute of Technology Madras, Chennai 600036, India.}

\date{\today}

\begin{document}

\maketitle

\begin{abstract}
  Baxter permutations are a class of permutations which are in bijection with a class of floorplans that arise in chip design called mosaic floorplans. We study 
  a subclass of mosaic floorplans called \texthfo{k} defined from mosaic floorplans by placing certain geometric restrictions. This 
  naturally leads to studying a subclass of Baxter permutations. This subclass 
  of Baxter permutations are characterized by pattern avoidance. We establish 
  a bijection, between the subclass of floorplans we study and a subclass of Baxter permutations,
  based on the analogy between decomposition of a floorplan into smaller blocks 
  and {\tmem{block}} decomposition of permutations. Apart from the 
  characterization, we also answer combinatorial
  questions on these classes. We give an algebraic generating function (but without a closed form solution) for the
  number of permutations, an exponential lower bound on growth
      rate, and a linear time algorithm for deciding membership in each subclass.
  Based on the recurrence relation describing the class, we also
  give a polynomial time algorithm for enumeration. We finally prove that
  Baxter permutations are closed under inverse based on an argument inspired from
  the geometry of the corresponding mosaic floorplans.
  This proof also establishes that the subclass of Baxter permutations 
  we study are also closed under inverse.
  Characterizing permutations instead of the 
  corresponding floorplans can be helpful in reasoning about the solution space
  and in designing efficient algorithms for floorplanning.
  \keywords{Floorplanning, Pattern Avoidance, Baxter Permutation}
\end{abstract}

\section{Introduction}

Baxter permutations are a well studied class of pattern avoiding permutations
having real world applications. One such application is to represent
floorplans in chip design. A floorplan is a rectangular dissection of a 
given rectangle into a finite number of indivisible rectangles using
axis parallel lines. These indivisible rectangles are locations in which 
modules of a chip can be placed. In the floorplanning phase of chip design, relative
positions of modules are decided so as to optimize cost functions like
wire length, routing, area etc. Given a set of modules and an associated cost
function, the floorplanning problem is to find an optimal floorplan. The
floorplanning problem for typical objective functions is
NP-hard~{\cite[p. 94]{sait1999vlsi}}. 
Hence combinatorial search algorithms
like simulated annealing~\cite{wong1986new} are used to find an optimal floorplan. The optimality
of the solution and performance of such algorithms depends on the class of
floorplans comprising the search space and their representation . Wong and
Liu~{\cite{wong1986new}} were the first to use combinatorial search for solving
floorplanning problems. They worked with a class of floorplans called slicing
floorplans which are obtained by recursively subdividing a given rectangle into two 
smaller rectangles either by a horizontal or a vertical cut. 
The slicing floorplans correspond to a class of permutations called separable
permutations~{\cite{Ackerman20061674}}. Later research in this direction
focused on characterizing and representing bigger classes of floorplans so
that search algorithms have bigger search spaces, potentially including the
optimum. One such category of floorplans is {\bf mosaic} floorplans which are a
generalization of slicing floorplans. Ackerman et al.~{\cite{Ackerman20061674}}
proved a bijection between mosaic floorplans and Baxter permutations. We study
a subclass of mosaic floorplans obtained by some natural restrictions on
mosaic floorplans. We use the bijection of Ackerman et al.~{\cite{Ackerman20061674}} as a tool to
characterize and answer important combinatorial problems related to this
class of floorplans. For the characterization of these classes we also use
characterization of a class of permutations called \tmtextit{simple}
permutations studied by Albert and Atkinson~{\cite{Albert20051}}.

Given a floorplan and dimensions of its basic rectangles, the area minimization
problem is to decide orientation of each cell which goes into basic rectangles
so as to minimize the total area of the resulting placement.
This problem is NP-hard for mosaic floorplans~\cite{STOCKMEYER198391}, but is polynomial time
for both slicing floorplans~\cite{STOCKMEYER198391} and \texthfo{5}~\cite{NET:NET10075}. 
Hence \texthfo{k} is an interesting class of floorplans with provably better performance
in area minimization~\cite{NET:NET10075} than mosaic floorplans. But the only representation
of such floorplans is through a top-down representation known as hierarchical tree~\cite{NET:NET10075} and is
known only for \texthfo{5}. Prior to this work it was not even known which floorplans
with $k$ rooms are non-sliceable and is not constructible hierarchically from 
mosaic floorplans of $k-1$-rooms or less. Such a characterization is needed to extend the
polynomial time area minimization algorithm based on non-dominance given in~\cite{NET:NET10075}.
We give such a characterization and provide an efficient representation for such floorplans by
generalizing generating trees to \textsgto{k}. We also give an exact characterization 
in terms of equivalent permutations.

Our main technical contributions are i) We establish a subclass of floorplans called
\texthfo{k}; ii) We characterize this
subclass of floorplans using a subclass of Baxter permutations; iii) We 
show that the subclass is exponential in size; iv) We present an algorithm 
to check the membership status of a permutation in the subclass of 
\texthfo{k}  and v) We present a simple proof of
closure under inverse operation for Baxter permutations 
using the mapping between the permutations and floorplans, and the geometry of the 
rectangular dissection. 

The remainder of the paper is organized as follows: in Section
\ref{sec:prelim}, we introduce the necessary background on floorplans and
pattern avoiding permutations. In Section \ref{sec:hfok}, we motivate and
characterize the subclasses of Baxter permutations studied in this paper. Section \ref{sec:combhfok} is devoted to answering interesting combinatorial
problems of growth, and giving generating function on these
subclasses. Section~\ref{sec:algorithmForMembership} gives an algorithm 
for membership in each class as well as for deciding given a Baxter permutation
the smallest $k$ for which it is \texthfo{k}.
Section \ref{sec:closurehfok} proves the closure
of Baxter permutations under inverse. Section \ref{sec:openproblems} lists some open problems. 
We also have a section  Appendix (see~\ref{app:figures}) 
which illustrates some floorplans which can be used to
gain intuition about the floorplan classes we define.

\section{Preliminaries}\label{sec:prelim}

\begin{figure}
\centering
\includegraphics[scale=0.2]{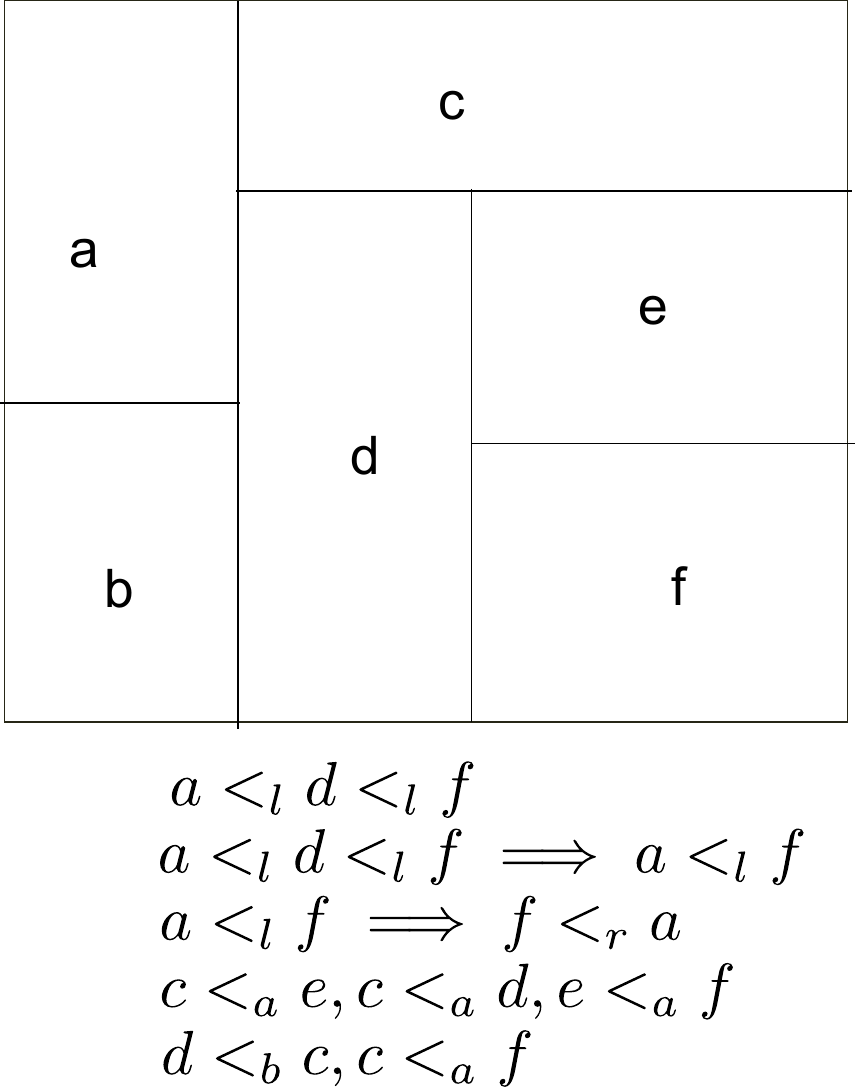}
\caption{ABLR relationships in a floorplan}
\label{fig:ablr}
\end{figure}

\begin{figure}
\centering
\includegraphics[scale=0.08]{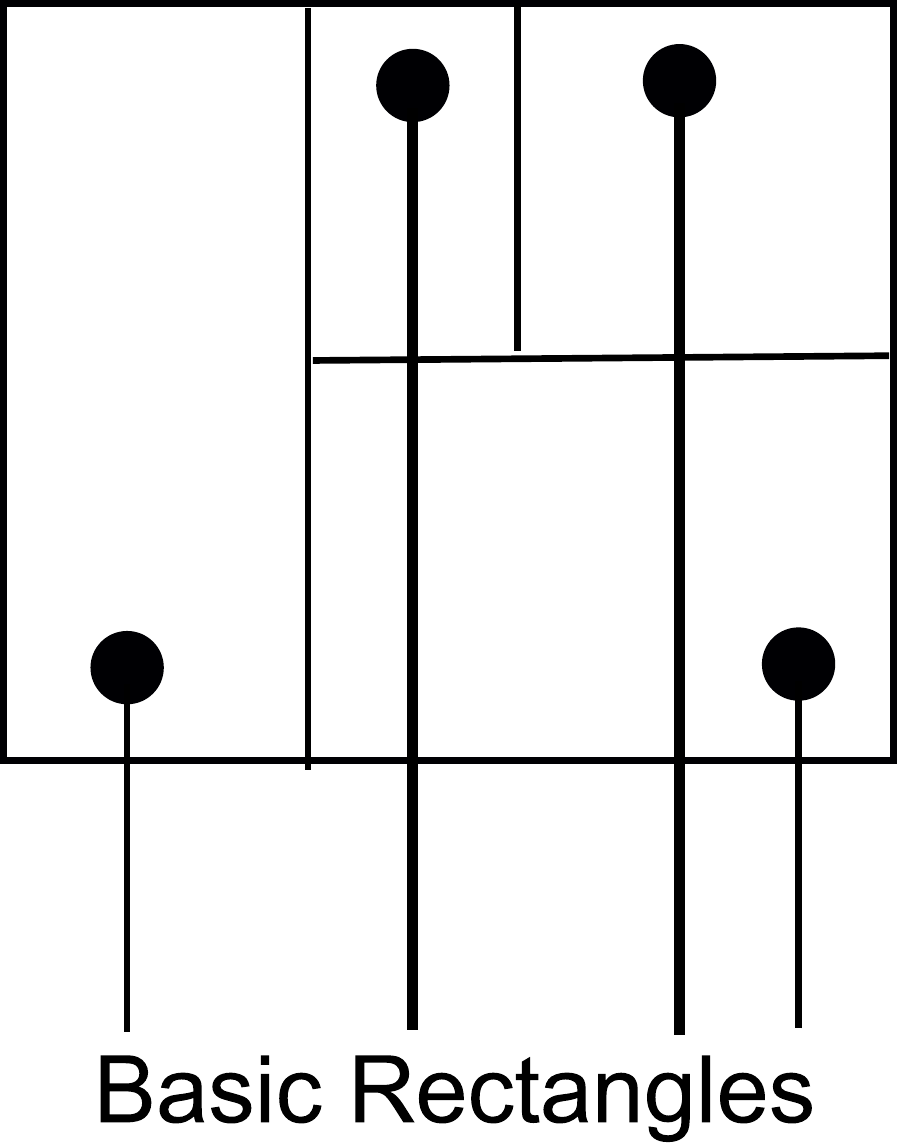}
\caption{A floorplan with its rooms marked}
\label{fig:rectangular_dissections}
\end{figure}

\begin{figure}
\centering
\includegraphics[scale=0.20]{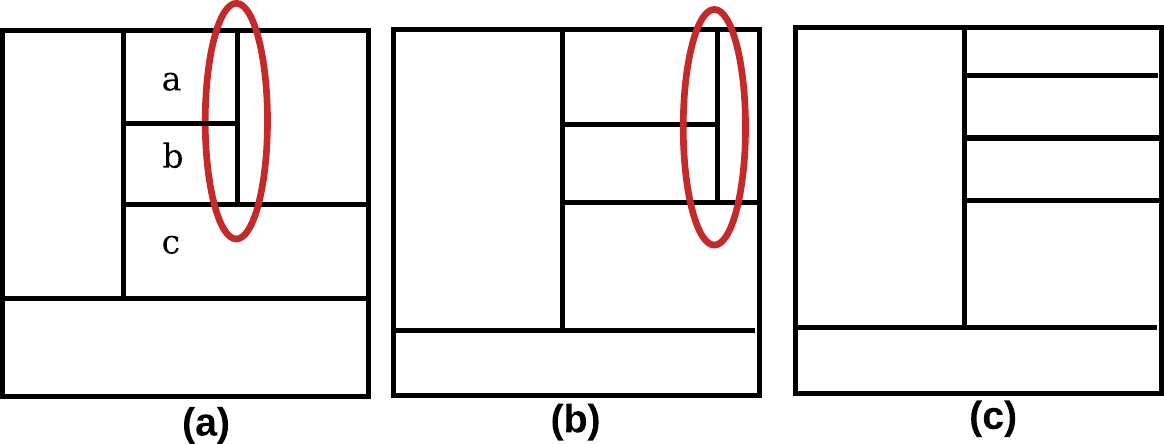}
\caption{Equivalence of Floorplans - $a \equiv b$, but $a \not\equiv c$}
\label{fig:equivalent_fp}
\end{figure}

A floorplan is a dissection of a given rectangle by line segments which are axis parallel (see Figure~\ref{fig:rectangular_dissections}).
The rectangles in a
floorplan which do not have any other rectangle inside are called basic rectangles or rooms.
For the remainder of the paper we will refer to them as rooms.
A
floorplan captures the relative position of the rooms via four relations
defined between rooms. Given a floorplan $f$, the ``left-of'' relation denoted by
$L_f$ is defined as $\left( a, b \right) \in L_f$ if there is a vertical line
segment of $f$ going through the right edge of room $a$ and left edge of room
$b$ or if there is a room $c$ such that $\left( a, c \right) \in L_f$ and
$\left( c, b \right) \in L_f$. 
When $(a,b) \in L_f$ we say that $a$ is to the ``left-of'' $b$ and is denoted
by $a <_l b$. 
For example in the floorplan given in Figure~\ref{fig:ablr} the room labeled $b$ is to the
left of room labeled $d$ because there is vertical segment through the right boundary of room $b$ and left boundary of room $d$.
Similarly for a floorplan $f$ the ``above'' relation
denoted by $A_f$ is defined as $\left( a, b \right) \in A_f$ if there is a
horizontal line segment of $f$ going through the bottom edge of room $a$ and
through the top edge of room $b$ or if there is a room $c$ such that $\left(
a, c \right) \in A_f$ and $\left( c, b \right) \in A_f$. The other two
relations are inverses of these relations: ``right-of'' is defined as $R_f =
\left\{ \left( a, b \right) \mid \left( b, a \right) \in L_f \right\}$ and
``below'' is defined as $B_f = \left\{ \left( a, b \right) \mid \left( b, a
\right) \in A_f \right\}$. A cross junction in a floorplan is an intersection
of two line segments such that the intersection point is not an end point of
either of the line segments. A mosaic floorplan is a floorplan where there are
no cross junctions.
This restriction is to ensure that, in a mosaic floorplan
between any two rooms, exactly one of $L_f, R_f, B_f, A_f$ holds~\cite[Observation~3.3]{Ackerman20061674}. 
 We denote the set of all mosaic floorplans with $k$ rooms by $M_k$.
The relations $X \in \{L_f,A_f,R_f,B_f\}$ can be naturally extended to 
that between rooms and line segments, by defining $(a,l) \in X$ if room $a$ is
supported by line segment $l$ from the respective direction $X$ in $f$.
We call two
mosaic floorplans $f_1, f_2$ equivalent if there is a bijective mapping $\psi : f_1
\rightarrow f_2$ \ such that $\left( a, b \right) \in X_{f_1}$ if and only if
$\left( \psi \left( a \right), \psi \left( b \right) \right) \in X_{f_2}$
where $X \in \left\{ L, R, A, B \right\}$ , i.e. $\psi$ preserves the relative
position of rooms and line segments. For example floorplans labeled $a,b$ in
Figure~\ref{fig:equivalent_fp} are equivalent under this definition whereas
$a$ and $c$ are not equivalent.

\begin{figure}
\centering
\includegraphics[scale=0.15]{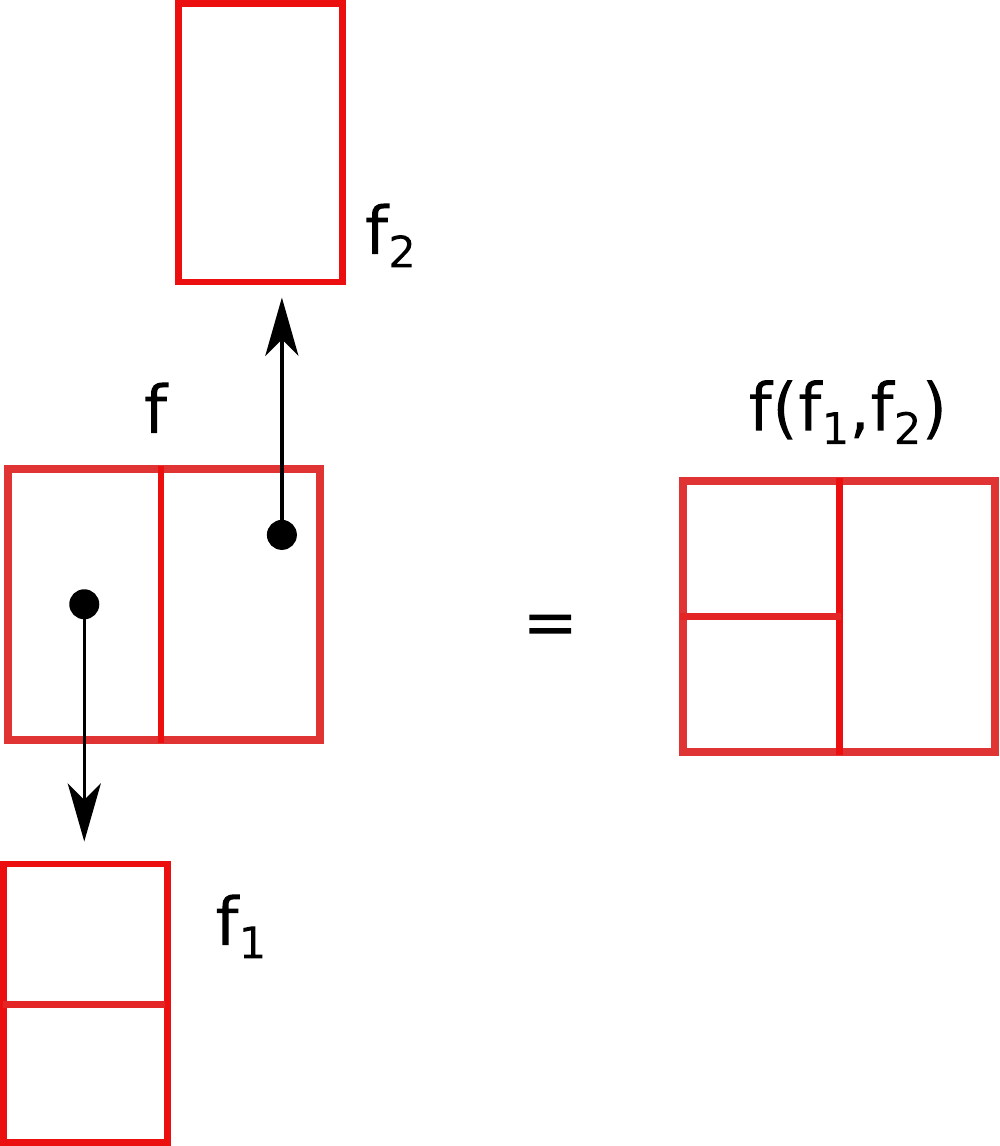}

\caption{Insertion operation on floorplans}
\label{fig:insertion_op}
\end{figure}

\begin{figure}
\centering
\includegraphics[scale=0.15]{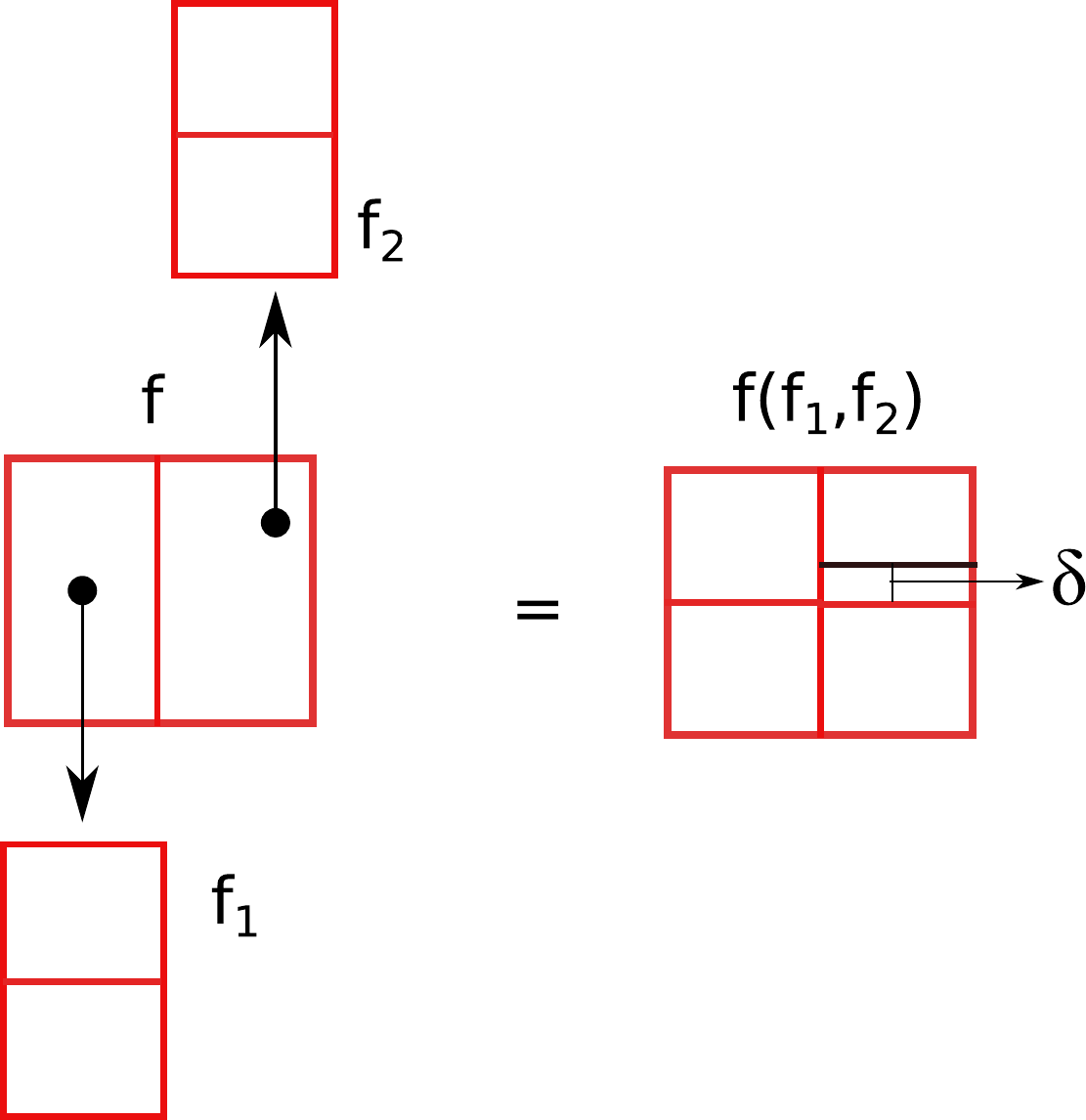}

\caption{Avoiding cross junction}
\label{fig:avoidcrossjn}
\end{figure}

In this paper we study a subclass of mosaic floorplans called \texthfo{k}.
The subclass \texthfo{k} for $k \geq 2, k \in
\mathbbm{N}$ (abbreviated as $\tmop{HFO}_k$ in the remainder of the paper) is
obtained by placing the following restriction on  mosaic floorplans : a mosaic
floorplan is $\tmop{HFO}_k$ if it can be constructed using mosaic floorplans with
at most $k$ rooms by repeated application of an operation which we call
{\tmem{insertion}}. 
\begin{definition}[Insertion]
Given a mosaic floorplan with $k$ rooms $f \in M_k$ and some fixed labeling of its rooms,
 {\tmem{insertion}} of $f$ by
$k$ mosaic floorplans $f_1, f_2, f_3, \ldots, f_k$ denoted by $f(f_1,\dots,f_k)$ is the mosaic floorplan
obtained by placing in $f_i$ in $i$th room  of $f$.
\end{definition}
Figure~\ref{fig:insertion_op} illustrates insertion of a floorplan with two rooms by
two other floorplans.
In insertion, if two adjacent rooms in $f$ (say $a$ and $b$) have
two segments coming
from inserted floorplans $f_a, f_b$ of same alignment (i.e., either both horizontal or both vertical)  
touching each other making a cross junction, then to make the resulting floorplan mosaic, one of the line
segments is moved by a small $\delta > 0$ as shown in Figure
\ref{fig:avoidcrossjn}. Moving a line segment by a small $\delta$ does not
change the relative position of rooms. This ensures that {\tmem{insertion}}
produces floorplans which are mosaic. 

We define a
mosaic floorplan $f$ to be decomposable if there exists $k > 1$ for which there
is a $g \in M_k$ and $k$ mosaic floorplans $g_1, \ldots, g_k$ at least one of which
is non trivial (i.e., has  more than one room) and $f=g(g_1,\dots,g_k)$.
A mosaic floorplan is
called in-decomposable if it is not decomposable.


\begin{figure}

\centering

\includegraphics[scale=0.2]{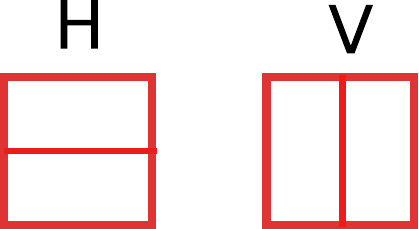}

\caption{$\tmop{HFO}_2$ building blocks}
\label{fig:hfovh}

\end{figure}


Ackerman et al.~{\cite{Ackerman20061674}} established a representation for
mosaic floorplans in terms of a class of pattern avoiding permutations called 
Baxter permutations. The bijection is 
established via two algorithms, one which produces a Baxter permutation
given a mosaic floorplan and another which produces a mosaic floorplan
given a Baxter permutation. For explaining the results in this paper we only need the algorithm 
which produces a Baxter permutation $\pi_f$ given a mosaic floorplan $f$. This algorithm
has two phases, a labeling phase where every room in the mosaic floorplan $f$ is
given a unique number in $\left[ n \right]$ and an extraction phase where the
labels of the rooms are read off in a specific order to form a permutation
$\pi_f \in S_n$. The labeling is done by successively removing the top-left room of current
floorplan by sliding it out of the boundary by pulling the edge which ends at
a T junction (since no cross junctions are allowed in a mosaic floorplan, for any room
every edge which is within the dissected rectangle is either a horizontal segment ending in a vertical segment
forming a $\dashv$ or is a horizontal segment on which a vertical segment ends
forming a $\perp$).
The $i$th floorplan to be removed in the above process is labeled room $i$ in
the original floorplan. 
After the labeling
phase we obtain a mosaic floorplan whose rooms are numbered from $[n]$. The permutation
corresponding to the floorplan is obtained in the second phase called
extraction where rooms from the bottom-left corner are successively removed by pulling the
edge ending at a T junction. The $i$th entry of the permutation $\pi_f$ is the
label of the $i$th room removed in the extraction phase.

\begin{figure}
\centering
\includegraphics[scale=0.2]{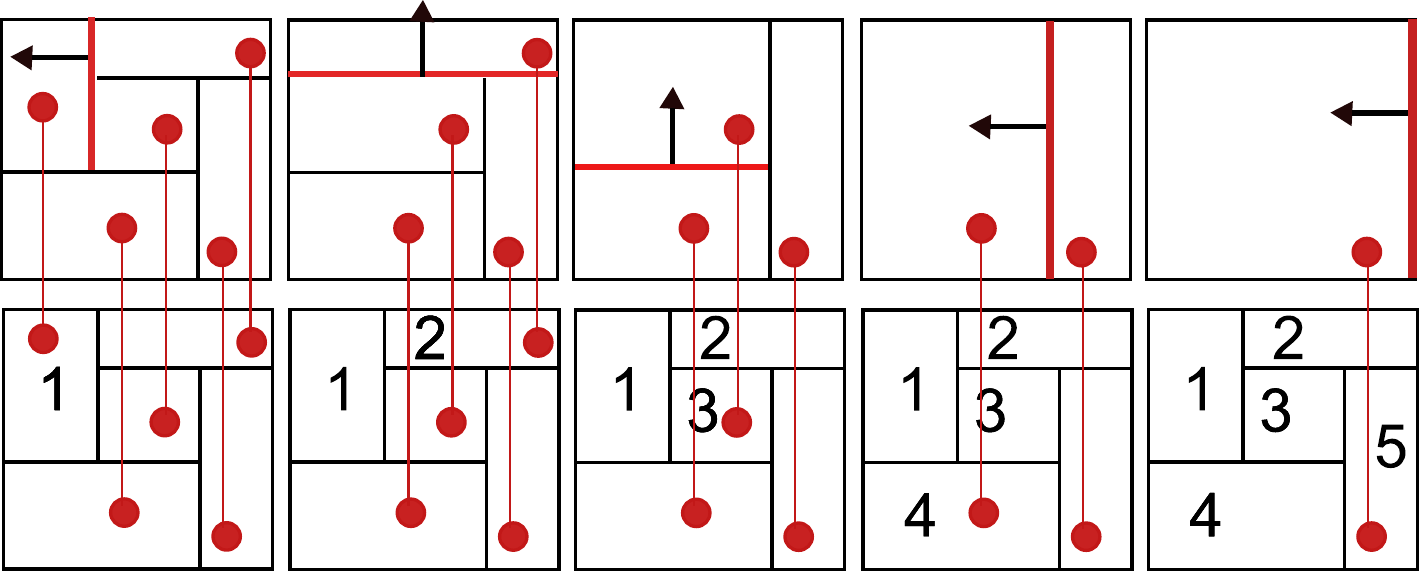}
\caption{FP2BP labeling phase}
\label{img:FP2BP_labelling}
\end{figure}

\begin{figure}
\centering
\includegraphics[scale=0.2]{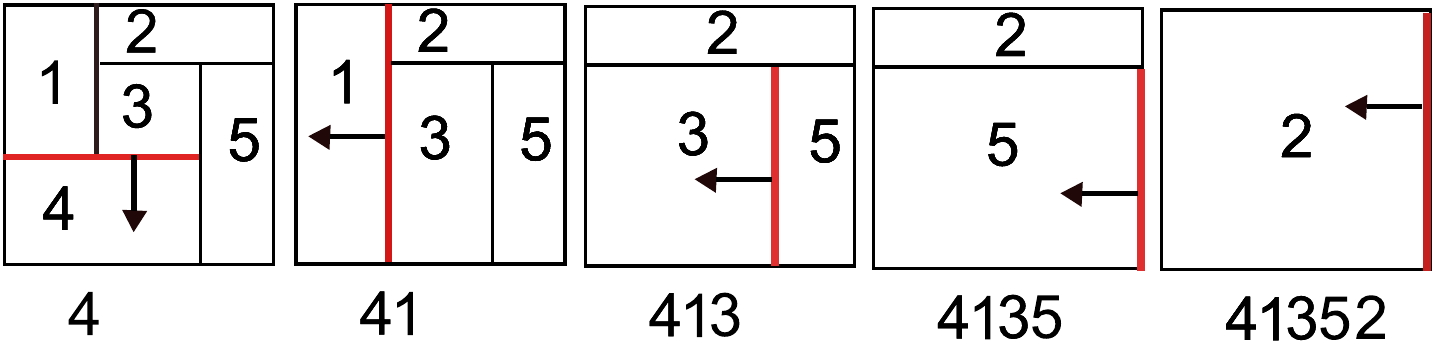}
\caption{FP2BP extraction phase}
\label{img:FP2BP_extraction}
\end{figure}

Figure \ref{img:FP2BP_labelling} demonstrates the labeling phase and Figure
\ref{img:FP2BP_extraction} demonstrates the extraction phase. If room $i$ 
is labeled before room $j$ then room $i$ is to the \tmtextit{left} or \tmtextit{above}
of room $j$, whereas if the room $i$ is removed before room $j$, i.e.,
$\pi^{- 1} \left[ i \right] < \pi^{- 1} \left[ j \right]$ then room $i$ is to
the \tmtextit{left} of or \tmtextit{below} room $j$ (see \cite[Observation 3.4]{Ackerman20061674}). Since the permutation
captures both the label and position of a room, it captures the
\tmtextit{above, below, left} or \tmtextit{right} relations between
rooms. Ackerman et al.(see \cite[Observation 3.5]{Ackerman20061674}) also proved that two rooms share an edge in a mosaic floorplan
$f$ if and only if either their labels are consecutive or their positions in
$\pi_f$ are consecutive.
For the rest of the paper we refer to this Algorithm of Ackerman et al. as $\tmop{FP2BP}$.

We now describe permutation classes which are used in this paper, including 
Baxter permutations mentioned earlier.
For the convenience of
defining pattern avoidance in permutations, we will assume that permutations
are given in the one-line notation (for ex., $\pi = 3142$). A permutation $\pi \in
S_n$ is said to contain a {\tmem{pattern}} $\sigma \in S_k$ if there are $k$
indices $i_1,\dots,i_k$ with $1 \leqslant i_1 < \cdots < i_k \leqslant n$ such that $\pi \left[ i_i
\right], \pi \left[ i_2 \right], \pi \left[ i_3 \right], \ldots, \pi \left[
i_k \right]$ called {\tmem{text}} has the same relative ordering as $\sigma$,
i.e., $\pi \left[ i_j \right] < \pi \left[ i_l \right]$ if and only if
$\sigma_j < \sigma_l$. 
Note that the sub-sequence need not be formed by consecutive entries in $\pi$.
If $\pi$ \tmtextit{contains} $\sigma$ it is denoted by $\sigma \leq \pi$.
A permutation $\pi$ {\tmem{avoids}} $\sigma$ if it does
not {\tmem{contain}} $\sigma$. For example $\pi = 4321$ {\tmem{avoids}}
$\sigma = 12$ because in $4321$ every number to the right of a
number is smaller than itself, but $\pi$ {\tmem{contains}} the {\tmem{pattern}}
$\rho = 21$ because numbers at any two indices of $\pi$ are in decreasing
order. A permutation $\pi$ is called \tmtextit{separable} if it
{\tmem{avoids}} the {\tmem{pattern}} $\sigma_1 = 3142$ and its reverse
$\sigma_2 = 2413$. Baxter permutations are a generalization of separable
permutations in the following sense: they are allowed to {\tmem{contain}}
$3142 / 2413$ as long as any $\pi \left[ i_1 \right], \pi \left[ i_2 \right],
\pi \left[ i_3 \right], \pi \left[ i_4 \right]$ which has the same relative
ordering as $3142 / 2413$ has $\left| \pi \left[ i_1 \right] - \pi \left[ i_4
\right] \right| > 1$. For example $\pi = 41532$ is not Baxter as {\tmem{text}}
$4153$ in $\pi$ matches {\tmem{pattern}} $3142$ and the absolute difference of
entry matching $3$ and entry matching $2$ is $4 - 3 = 1$. However $\pi =
41352$ is a Baxter permutation as the only text which matches $3142$ is $4152$ and the
absolute difference of entries matching $3, 2$ is $4 - 2 = 2$ which is greater
than $1$.

Another class of permutations important to this study is the class of
{\tmem{simple}} permutations. They are a class of block in-decomposable
permutations. To define this in-decomposability we need the following
definition : a {\tmem{block}} of a permutation is a set of consecutive
positions such that the values from these positions form an interval $\left[
i, j \right]$ of $\mathbb{N}$. Note that the values in the block need not be 
in ascending order as it is in the interval corresponding to the block $[i,j]$.
The notion of block in-decomposability is
defined by a decomposition operation called {\tmem{inflation}}. We
recall the definition from Section 2 of~\cite{Albert20051}.
\begin{definition}[Inflation]
Given a permutation $\sigma \in S_k$, {\tmem{inflation}} of $\sigma$
by $k$ permutations $\rho_1, \rho_2,
\rho_3, \dots, \rho_k$, denoted by $\sigma \left( \rho_1,
\ldots, \rho_k \right)$ is the permutation $\pi$ where each element $\sigma_i$
of $\sigma$ is replaced
with a {\tmem{block}} of length $|\rho_i|$ whose elements have the same relative
ordering as $\rho_i$, and the blocks
among themselves have the same relative ordering as $\sigma$.
\end{definition}

For example
inflation of $3124$ by $21, 123, 1$ and $12$ results in $\pi = 65 \, 123 \, 4 \,
78$ where $65$ is the {\tmem{block}} corresponding to $21$, $123$ corresponds
to $123$, $4$ corresponds to $1$ and $78$ corresponds to $12$. If $\pi =
\sigma \left( \rho_1, \ldots, \rho_k \right)$ then $\sigma \left( \rho_1,
\ldots, \rho_k \right)$ is called a \tmtextit{block-decomposition} of $\pi$. A
\tmtextit{block-decomposition} \ $\sigma \left( \rho_1, \rho_2, \ldots, \rho_k
\right)$ is non-trivial if $\sigma \in S_k$ for $k > 1$ and at least one
$\rho_i$ is a non-singleton permutation (i.e. of more than one element).
A permutation is \tmtextit{block-in-decomposable} if it has no non-trivial
block-decomposition.
Note that {\tmem{inflation}} on permutations as defined above is analogous to
{\tmem{insertion}} on mosaic floorplans defined earlier.

Block in-decomposable permutations can be thought of as building blocks of all
other permutations by \tmtextit{inflations}. Albert and Atkinson
{\cite{Albert20051}} studied \tmtextit{simple} permutations
which are permutations whose only \tmtextit{blocks} are the trivial
\tmtextit{blocks} (which is either a single point $\pi \left[ i \right]$ or the
whole permutation $\pi \left[ 1 \ldots n \right]$). They also defined a sub
class of \tmtextit{simple} permutations called \tmtextit{exceptionally simple}
permutations which are defined based on an operation called
\tmtextit{one-point deletion}. A one-point deletion on a permutation $\pi \in
S_n$ is deletion of a single element at some index $i$ and getting a new
permutation $\pi' \in S_{n - 1}$ by rank ordering the remaining elements. For
example one-point deletion at index $5$ of $41352$ gives $4135$ which when
rank ordered gives the permutation $3124$. A permutation $\pi$ is
\tmtextit{exceptionally simple} if it is \tmtextit{simple} and no
\tmtextit{one-point deletion} of $\pi$ yields a simple permutation. Albert and Atkinson~{\cite{Albert20051}}
characterized \tmtextit{exceptionally simple} permutations and proved that for
any permutation $\pi \in S_n$ which is \textit{exceptionally simple} there exists two successive
\tmtextit{one-point} deletions which yields a \tmtextit{simple} permutation
$\pi' \in S_{n - 2}$.

\section{Characterizing \texthfo{k}}\label{sec:hfok}
In this section we characterize \texthfo{k}
in terms of corresponding permutations using the notion of block decomposition
defined earlier.

We note that this connection can be seen for a level of the hierarchy
well studied in literature, namely $\tmop{HFO}_2$.
$\tmop{HFO}_2$, the class of floorplans which can be built by repeated
application of insertion of the two basic floorplans shown in Figure
\ref{fig:hfovh} are also called slicing floorplans. Slicing floorplans are known~{\cite{Ackerman20061674}} to
be in bijective correspondence with {\tmem{separable permutations}}. Separable
permutations are also the class of permutations $\pi$ such that it can be
obtained repeated {\tmem{inflation}} of $1$ (the singleton permutation) by,
$12$ or $21$. Note that both $12, 21$ are {\tmem{simple}} permutations. Even
though $\tmop{HFO}_2$ is well studied in literature and is known to be in
bijective correspondence with separable permutations, the connection to block
decomposition of permutations was not explicitly observed.

\begin{figure}[h]

\centering

\includegraphics[scale=0.2]{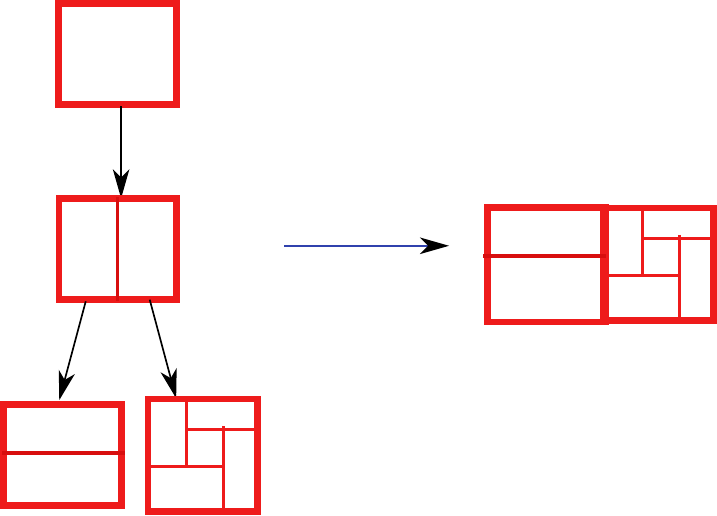}

\caption{Generating tree corresponding to an $\tmop{HFO}_5$ floorplan}
\label{img:generatingtreehfo5}

\end{figure}

\begin{figure}

\centering

\includegraphics[scale=0.2]{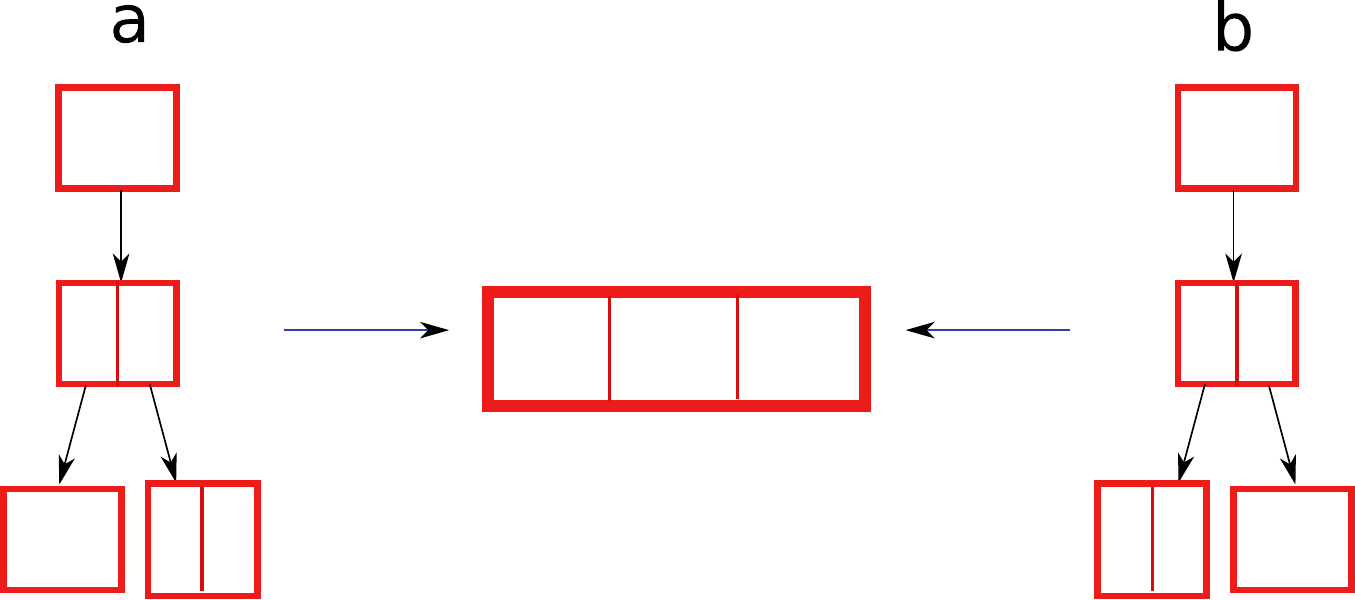}

\caption{Ambiguity in vertical cuts}
\label{img:generatingtreeamguity}

\end{figure}

$\tmop{HFO}_5$ (shown in Figure~\ref{fig:wheels} and Figure~\ref{fig:hfofiveninerooms})
floorplans are also studied in the literature, but the
only characterization till date for these floorplans is based on a discrete
structure called {\tmem{generating trees}}. We generalize this structure for
an arbitrary $k$ in the following sense : a {\tmem{generating tree}} of order
$k$ is a rooted tree, where each node is labeled by an in-decomposable mosaic
floorplan, say $g$ of at most $k$ rooms, and the number of children of a node is equal
to the number of rooms in the floorplan labeling the node. 
The children are arranged in the order $\pi^{-1}_g$ from left to right. 
That is the left most child corresponds to the first room to be removed in the 
extraction phase of $\tmop{FP2BP}$ and second from left corresponds to second room 
to be removed and so on and so forth.
The
{\tmem{generating tree}} captures the top down application of
{\tmem{insertion}}'s to yielding the given floorplan in the following sense : an
internal node of a {\tmem{generating tree}} represents insertion of $f$ - the
floorplan labeling the node - by the floorplans labeling its children, $f_1,
\ldots, f_k$ (ordered from left to right).
Figure \ref{img:generatingtreehfo5} is a generating tree for an
$\tmop{HFO}_5$ floorplan. There could be more than one generating tree for a
floorplan owing to the fact there is ambiguity in consecutive vertical slices
and in consecutive horizontal slices, as illustrated in Figure
\ref{img:generatingtreeamguity}. But this can be removed (proved later) by
introducing two disambiguation rules called ``skew''. Skew rule insists that
when there are multiple parallel vertical (respectively, horizontal) line segments
touching the bounding box of the floorplan $f$, we consider only the insertion
operation $f_1, f_2$ where $f_2$ is the floorplan contained to the right of (respectively, above)
the first parallel line segment from left (respectively, bottom) and $f_1$ is the floorplan
contained to the left of (respectively below) the first parallel line segment from left (respectively, bottom).
Hence
only the tree labeled $a$ satisfies ``skew'' rule among the generating trees
in Figure~\ref{img:generatingtreeamguity}. A generating tree satisfying
``skew'' rule is called \tmtextit{Skewed Generating Tree}.




The connection between {\tmem{insertion}} and block decomposition and the fact
the bijection of Ackerman et al.{\cite{Ackerman20061674}} preserves this
connection is the central idea of our paper. 
The following observation about the algorithm FP2BP, though not mentioned in
the original paper, is not hard to see, but is useful for the characterization
of $\tmop{HFO}_k$.

\begin{lemma}
  For a mosaic floorplan $f$ let $\pi_{f}$ denote the unique
  Baxter permutation obtained by algorithm $\tmop{FP2BP}$. If
  $f=g(g_1,\dots,g_k)$  i.e., it is obtained by 
  insertion of $g \in M_k$ by $g_1, \ldots, g_k$, then
  \[ \pi_f = \pi_g \left( \pi_{g_1}, \ldots, \pi_{g_k} \right) \]
  where $\pi_g \left( \pi_{g_1}, \ldots, \pi_{g_k} \right)$ denotes the
  permutation obtained by inflating $\pi_g$ with $\pi_{g_1}, \ldots,
  \pi_{g_k}$.\label{lemma:FPMosaicInjection}
\end{lemma}

\begin{proof}
  Since $f$ is obtained by insertion of $g \in M_k$ by $g_1, \ldots, g_k$,
  each $g_i$ is completely contained inside a rectangle, the $i$th room
  of $g$. The theorem follows from the fact that FP2BP labeling labels
  all the rooms contained inside a rectangle before moving out, and it
  extracts all the rooms inside a rectangle before moving out of the rectangle. 
  
  We will first prove that the FP2BP labeling labels
  all the rooms contained inside a rectangle before moving out.
  To prove this assume to the contrary that there exists rooms $a, b, c$ with $a$ and
  $b$ belonging to $g_i$ and $c$ belonging to $g_j, j \neq i$ such that they are
  labeled in the order $a, c, b$ without loss of generality. 
  By the property (see \cite[Observation~3.4]{Ackerman20061674}) of the labeling algorithm
  $a$ is to the left or above of $c$, and $c$ is to the left or above $b$ and
  since they are labeled consecutively there is a line segment shared by $a$
  and $c$ as well as $c$ and $b$. They can only be oriented in one
  of the four ways shown in Figure~\ref{img:orientationsLabeling} corresponding to
  whether $a <_l c$ or $a <_a c$ and $c <_l b$ or $ c <_a b$. Among the four, 
  except for $a <_l c <_a b$ and its symmetric counterpart $a <_a c <_l b$,
  it is clear that it cannot be the case that $a$ and $b$ are contained in
  one rectangle but $c$ in another. For the orientation $a  <_l c <_a b$, 
  the fact that there is a line segment shared by $b$ and $c$
  removes the possibility of $a, b$ being in one rectangle and $c$ being in
  another. 
  
  A symmetric argument can be used to establish the same when $a <_a c <_l b$.
  A similar argument can be used to establish that the extraction algorithm 
  moves to another rectangle only after exhausting all the rooms in the current rectangle.

\begin{figure}

\centering

\includegraphics[scale=0.3]{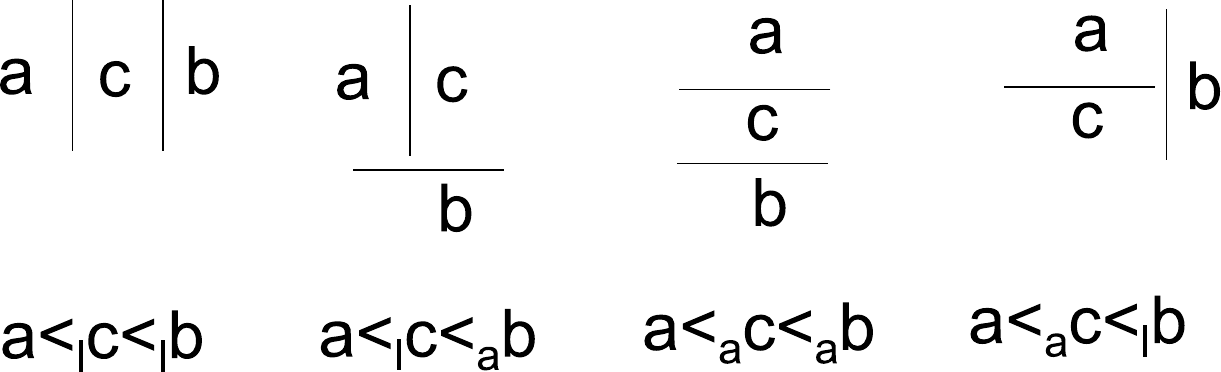}

\caption{Possible orientations of blocks $a, c, b$ labeled in that order}
\label{img:orientationsLabeling}

\end{figure}

\end{proof}

We obtain the following useful corollary from Lemma~\ref{lemma:FPMosaicInjection} (see Appendix for a proof~\ref{app-coro:mosaicindecomp}):

\begin{corollary}
  A mosaic floorplan $f$ is in-decomposable if and only if the Baxter
  permutation $\pi_f$ corresponding to it is block
  in-decomposable.\label{coro:mosaicindecomp}
\end{corollary}

\begin{figure}

\centering

\includegraphics[scale=0.2]{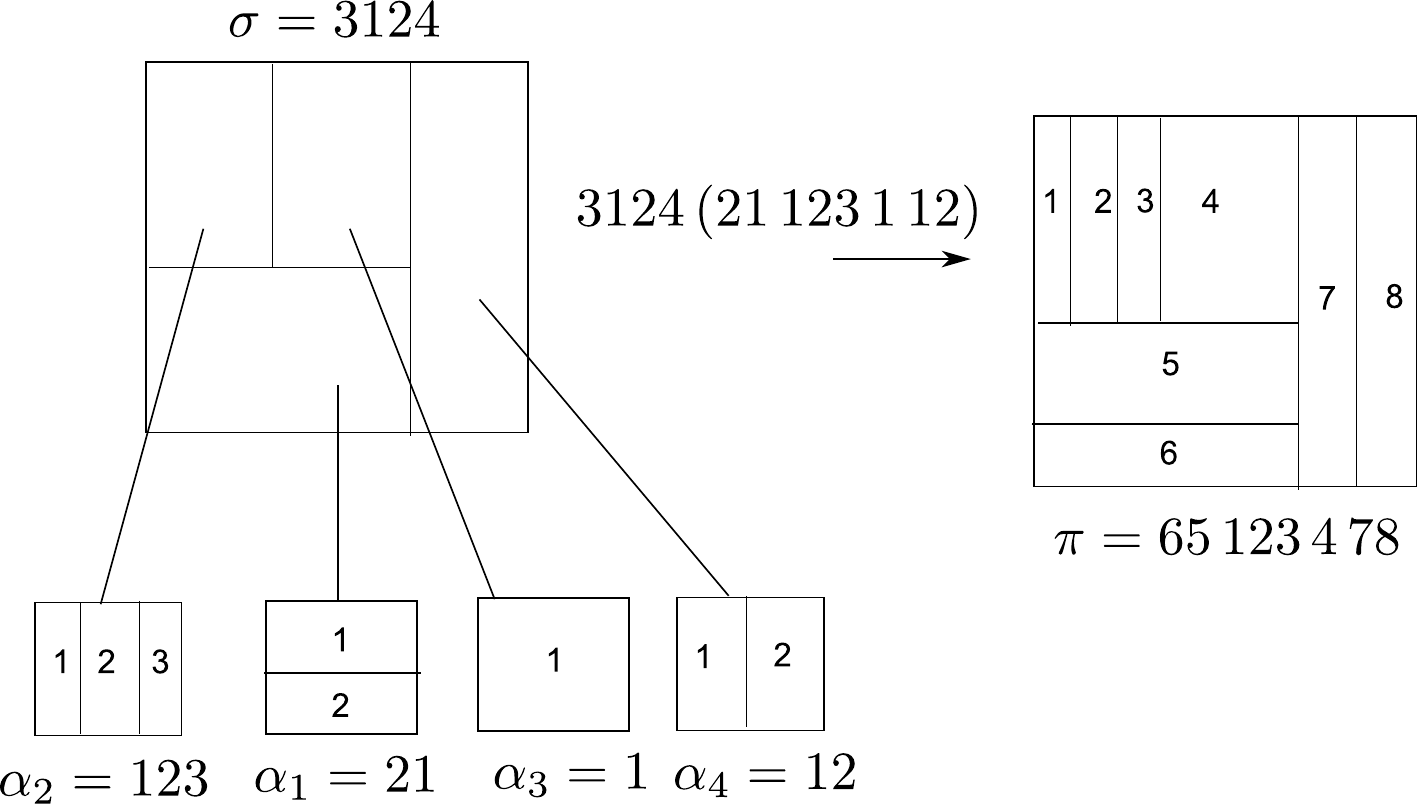}

\caption{Correspondence between inflation and insertion}
\label{fig:corrinfinj}

\end{figure}

For the characterization we will also need the following connection between
generating trees and block decomposition of permutations. Let $T_f$ be a
generating tree corresponding to $f$, satisfying the ``skew'' rule, then $T_f$
captures the unique block decomposition of a permutation as defined in~\cite[Proposition~2]{Albert20051}.
Label every node of $T_f$ by Baxter
permutation $\pi_{f_i}$ corresponding to the mosaic floorplan $f_i$ labeling
it. Mosaic floorplan $g$ corresponding to the sub-tree rooted at $f_i$ is
obtained by the insertion of $f_i$ by the floorplans labeling its children
$f_{i_1}, \ldots, f_{i_k}$. Hence by applying Lemma
\ref{lemma:FPMosaicInjection} we get that $\pi_g = \pi_{f_i} \left(
\pi_{f_{i_1}}, \ldots, \pi_{f_{i_k}} \right)$. So generating trees labeled by
Baxter permutations $\pi_{f_i}$ captures the block decomposition of Baxter
permutation $\pi_f$ corresponding to the floorplan $f$. Figure
\ref{fig:corrinfinj} illustrates the correspondence between
\tmtextit{inflation} and \tmtextit{insertion} by showing the equivalence
between \tmtextit{inflating} $3124$ with $123, 21, 1$ and $24$, and
\tmtextit{inserting} the floorplan corresponding to $3124$ with floorplans
corresponding to $123, 21, 1$ and $24$.

\begin{theorem}\label{thm:generating_tree_correspondence}
  \textsgto{k} are in bijective correspondence with
  $\tmop{HFO}_k$ floorplans. Moreover they capture the block decomposition of
  the Baxter permutation corresponding to the floorplan.\label{thm:skwdgt}
\end{theorem}

\begin{proof}
  It follows from definition of $\tmop{HFO}_k$ that there is a
  \tmtextit{generating tree of order $k$} capturing the successive
  applications of insertions resulting in the final floorplan. Since
  $\tmop{HFO}_k$ are a subclass of mosaic floorplans which are in
  bijective correspondence with Baxter permutations, there is unique Baxter
  permutation $\pi_f$ corresponding to the floorplan $f$. Lemma
  \ref{lemma:FPMosaicInjection} can now be used to prove that a
  \tmtextit{generating tree of order $k$} captures the block decomposition of
  $\pi_f$, by induction on the height of the tree. Consider the base case to be
  $h = 1$, i.e, the whole tree is one node labeled by an in-decomposable
  mosaic floorplan $f$ and by Corollary \ref{coro:mosaicindecomp}, $\pi_f$ is
  block in-decomposable. Assume that for any $h < l$, \tmtextit{generating
  trees of order} $k$ captures the block decomposition of $\pi_f$. Take a tree
  of height $h = l$ corresponding to a floorplan $f$, and let the root node be
  labeled by $g$ and children be labeled $g_1, \ldots, g_k$. By Lemma
  \ref{lemma:FPMosaicInjection}, $\pi_f = \pi_g \left( \pi_{g_1}, \ldots,
  \pi_{g_k} \right)$. We can apply induction hypothesis on the children to get
  the decomposition of $\pi_{g_1}, \ldots, \pi_{g_k}$.
  
  To prove the uniqueness of skewed generating trees we use the following
  theorem by Albert and Atkinson~\cite[Proposition~2]{Albert20051}
  proving the uniqueness of the block-decomposition represented by skewed
  generating trees.
  
  \begin{theorem}
    For every non singleton permutation $\pi$ there exists a unique simple non
    singleton permutation $\sigma$ and permutations $\alpha_1, \ldots,
    \alpha_n$ such that
    \[ \pi = \sigma \left( \alpha_1, \ldots, \alpha_n \right) \]
    Moreover if $\sigma \neq 12, 21$ then $\alpha_1, \ldots, \alpha_n$ are
    also uniquely determined. If $\sigma = 12$ (respectively, $21$) then
    $\alpha_1$ and $\alpha_2$ are also uniquely determined subject to the
    additional condition that $\alpha_1$ cannot be written as $\left( 12
    \right) \left[ \beta, \gamma \right]$ (respectively as $\left( 21 \right)
    \left[ \beta, \gamma \right]$)
  \end{theorem}
  
  The proof is completed by noting that the decomposition obtained by \textsgto{k}
   satisfies the properties of the
  decomposition described in the above theorem. 
  In a skewed generating tree if parent is $\sigma = 12$(respectively, $21$), 
  then its left child cannot be $12$(respectively, $21$). 
  Hence the block-decomposition corresponding to the
  left child, $\alpha_1$, cannot be $\left( 12 \right) \left[ \beta, \gamma
  \right]$ (respectively, $\left( 21 \right) \left[ \beta, \gamma \right]$).
  Since such a decomposition is unique, the skewed generating tree also must be unique.
  Hence the theorem. 
\end{proof}

To characterize $\tmop{HFO}_k$ in terms of pattern avoiding permutations
the following insight is used: if a
permutation $\pi$ is Baxter then it corresponds to a mosaic floorplan. Every
mosaic floorplan is $\tmop{HFO}_k$ for some $k$. Hence for a Baxter permutation $\pi$
the corresponding floorplan $f_\pi$ is not
$\tmop{HFO}_k$ for some specific $k$, it will be because of existence of a
node  in the unique skewed generating tree
corresponding to $f_{\pi}$, which is labeled by an in-decomposable mosaic floorplan 
$g \in \tmop{HFO}_l$ for some $l > k$. Since $\pi$ is obtained by inflation of
permutations including $\pi_g$ corresponding to $g$, $\pi$ will have some text
which matches the \tmtextit{pattern} $\pi_g$ because of the Lemma
\ref{lemma:blockdecomposition}. Thus if we can figure out all the
\tmtextit{patterns} which correspond to in-decomposable mosaic floorplans
which are $\tmop{HFO}_l$ for some $l > k$ then $\tmop{HFO}_k$ would be all
Baxter permutations which avoid those \tmtextit{patterns}. We defer the proof of 
Lemma~\ref{lemma:blockdecomposition} to the Appendix (see~\ref{app-lemma:blockdecomposition}).
\begin{lemma}
  If $\pi = \sigma \left( \rho_1, \ldots, \rho_k \right)$, then $\pi$ contains
  all patterns which any of $\sigma, \rho_1, \rho_2, \ldots, \rho_k$
  contains.\label{lemma:blockdecomposition} 
\end{lemma}

  We will use the following lemma which is proved in the Appendix (see~\ref{app-lemma:indecomposable_implies_inflation}).

\begin{lemma}
  If $\pi = \sigma \left( \rho_1, \ldots, \rho_k \right)$, then any 
  block in-decomposable pattern in $\pi$ has a matching text which is completely 
  contained in one of $\sigma,\rho_1,\rho_2,\ldots,\rho_k$.
  \label{lemma:indecomposable_implies_inflation}
\end{lemma}

Let $f$ be an in-decomposable mosaic floorplan which is $\tmop{HFO}_l$ for
some fixed $l \in \mathbb{N}$. By Corollary~\ref{coro:mosaicindecomp}, the
permutation corresponding to $f$, $\pi_f$ would be block in-decomposable and
hence it will be a \tmtextit{simple} permutation of length $l$. It is known (see \cite[Theorem 5]{Albert20051}) that a simple permutation of length
$l$ has either a \tmtextit{one-point} deletion which yields another simple
permutation or two \tmtextit{one-point} deletions giving a simple
permutation. Hence by successive applications of \tmtextit{one-point}
deletions we can reduce $\pi_f$ to a \tmtextit{simple} permutation of length
$k$, or an \tmtextit{exceptionally simple} permutation of length $k + 1$ (at
which point there is no further one \tmtextit{one-point} deletion giving a
simple permutation) for any $k < l$. Also if $\pi'$ is obtained from $\pi$ by
a \tmtextit{one-point} deletion at index $i$, then $\pi \left[ 1, \ldots,
i - 1, i + 1, \ldots, n \right]$ matches the \tmtextit{pattern} $\pi'$. 
That is $\pi$ contains all patterns $\pi'$ which are permutations obtained by
one point deletion of $\pi$ at some index. Also since pattern containment is
transitive by definition, if $\pi''$ is obtained by one-point deletion of $\pi'$
which in turn obtained from $\pi$ by a one-point deletion, then $\pi'' \leq \pi'$ and $\pi' \leq \pi$ implies that  $ \pi'' \leq \pi$.
From
these observations we get the following characterization of $\tmop{HFO}_k$.

\begin{theorem}\label{thm:hfokcharacterization}
  A mosaic floorplan $f$ is $\tmop{HFO}_k$ if and only if the permutation
  $\pi_f$ corresponding to $f$ (obtained by algorithm $\tmop{FP2BP}$)
  does not contain patterns from simple permutations of length $k + 1$ or
  exceptionally simple permutations of length $k + 2$.
\end{theorem}

\begin{proof}
  By Theorem~\ref{thm:skwdgt}, for any $\tmop{HFO}_k$ floorplan $f$
  there is a unique \textsgto{k}, $T_f$ such that
  it captures the block-decomposition of $\pi_f$. And in the block-decomposition of
  a generating tree of order $k$, permutations corresponding to
  the nodes are labeled by $\tmop{HFO}_k$ permutations of length at most $k$.
  Hence the block-decomposition of $\pi_f$ contains only block in-decomposable 
  permutations of length at most $k$. By Lemma~\ref{lemma:indecomposable_implies_inflation}
  $\pi_f$ cannot contain patterns which are block in-decomposable permutations of length 
  strictly more than $k$. Thus $\pi_f$ cannot contain patterns from simple
  permutations of length $k+1$ or from exceptionally simple permutations of length $k+2$
  as they are both classes of block in-decomposable permutations of length strictly greater than $k$.
  
  For the reverse direction, we prove that any mosaic floorplan which is $\tmop{HFO}_l, l > k$
  contains either a simple permutation of length $k+1$ or an exceptionally
  simple permutation of length $k+2$. From the fact that by definition any mosaic floorplan is
  $\tmop{HFO}_j$ for some $j$ and the forward direction that 
  no $\tmop{HFO}_k$ floorplan contains either a simple permutation of length $k+1$
  or an exceptionally simple permutation of length $k+2$ proof is completed.
  Suppose if it is $\tmop{HFO}_l$ for $l > 0$ then $\pi_f$ would have a
  \tmtextit{text} matching a pattern $\sigma \in S_l$ which is a simple
  permutation. Because the generating tree $T_f$ will have $\sigma$ and so would
  the block decomposition of the sub-tree rooted at node $\sigma$. And by Lemma
  \ref{lemma:blockdecomposition}, $\pi_f$ would also contain $\sigma$. 
  From $\sigma$ we can obtain 
  by successive \tmtextit{one-point} deletions a permutation $\sigma'$ which is either a simple permutation of length
  $k$ or is an exceptionally simple permutation of length $k + 1$.
  And $\sigma'$ would
  match a \tmtextit{text} in $\pi_f$ because $\pi_f$ had a \tmtextit{text}
  matching $\sigma$ and $\sigma$ contains this permutation, i.e., $\sigma'' \leq \sigma \leq \pi_f \implies \sigma'' \leq \pi_f$.
  
\end{proof}

From the above characterization it can be proved that the hierarchy
$\tmop{HFO}_k$ (it is a hierarchy because by definition $\tmop{HFO}_i \subseteq
\tmop{HFO}_{i + 1}$) is strict for $k \geqslant 7$, i.e. there is at least one
floorplan which is $\tmop{HFO}_k$ but is not $\tmop{HFO}_i$ for any $i < k$.
The natural candidates for such separation are in-decomposable mosaic
floorplans on $k$ rooms which corresponds to \tmtextit{simple} permutations of
length $k$ which are Baxter. It is easy to verify that for $k = 5$, $\pi_5 =
41352$ is such a permutation. Note that $\pi_5$ is of the form $\pi \left[ n -
1 \right] = n$ and $\pi \left[ n \right] = 2$. From $\pi_5$ we can obtain
$\pi_7 = 6413572$ by inserting $7$ between $5$ and $2$ and appending $6$ at
the beginning. It can be verified that $\pi_7$ is not $\tmop{HFO}_5$. It turns
out that all permutations of length at most $4$ which are Baxter are also
$\tmop{HFO}_2$, making $\tmop{HFO}_5$ the first odd number from where one can
prove the strictness of the hierarchy. Also every $\tmop{HFO}_6$ is
$\tmop{HFO}_5$, hence for even numbers separation theorem can
only start from $8$. Hence we prove the separation theorem for $k \geq 7$ 
generalizing the earlier stated idea. The generalization builds a $\pi_{k + 2}$ from a
  $\pi_k$ which is an in-decomposable $\tmop{HFO}_k$ having $\pi \left[ n - 1
  \right] = n$ and $\pi \left[ n \right] = 2$, by setting $\pi_{k + 2} \left[
  1 \right] = n + 1$, $\pi_{k + 2} \left[ i \right] = \pi_k \left[ i - 1
  \right], 2 \leqslant i \leqslant n$, $\pi_{k + 2} \left[ n + 1 \right] = n +
  2$ and $\pi_{k + 2} \left[ n + 2 \right] = 2$. The proof of the theorem is
  deferred to the Appendix (see ~\ref{app-thm:hfokseparataion}).
  \begin{theorem}
  For any $k \geq 7$, there exists a floorplan $f$ which is in $\tmop{HFO}_{k
  + 2}$ but is not in $\tmop{HFO}_{l}$ for any $l \leq k+1$
  \label{thm:hfokseparataion}
\end{theorem}

\section{Combinatorial study of $\tmop{HFO}_k$}\label{sec:combhfok}

We will first prove for any fixed $k$ the existence of a rational generating function for
$\tmop{HFO}_k$. Since we have proved that the number of
distinct $\tmop{HFO}_k$ floorplans with $n$ rooms is equal to the number of
distinct \textsgto{k} with $n$ leaves, it suffices to
count such trees. Let $t^k_n$ denote the number of distinct \textsgto{k}
 with $n$ leaves and $t^k_1$ represent a
rectangle for any $k$. Hence to provide a rational generating function for
number of distinct $\tmop{HFO}_k$ floorplans with $n$ rooms,
it suffices to provide one for the count $t^k_n$.

We will first describe the method for $\tmop{HFO}_5$. For 
simplicity of analysis let $t_i=t_i^5$.
 \textsgto{5} are
labeled by simple permutation of length at most $5$ which are
Baxter. There are only four of them - $12,21,25314$ and $41352$. Thus
the root node of such a tree also must be labeled from one these four
permutations. We obtain a recurrence by partitioning the set of
\textsgto{5} into four classes decided by the label of the root.  Let
$a_n$ denote the number of \textsgto{5} with $n$ leaves whose root is labeled $12$,
$b_n$ denote the number of \textsgto{5} with $n$ leaves whose root is labeled $21$,
$c_n$ denote the number of \textsgto{5} with $n$ leaves whose root is labeled
$41352$ and $d_n$ the number denote the \textsgto{5} with $n$ leaves whose root
is labeled $25314$. Since these are the only in-decomposable
$\tmop{HFO}_k$ permutations for $k \leqslant 5$, the root (and also
any internal node) has to labeled by one of these permutations. Hence
we get the following recurrence for $t^5_n$,
$t_n^5 = a_n + b_n + c_n + d_n$.

In a skewed tree if the root is labeled $12$, its left child cannot be $12$
but it can be $21$, $41352$ ,$25314$ or a leaf node. 
Hence the left child of the root of a tree in $a_n$ has to be labeled from
$b,c$ or $d$, but the right child has no such restriction. By definition
of skewed generating trees if the root
is labeled by a permutation of length $l$, it will have $l$ children, such that
the number of leaves of the children sum to $n$.
Hence if root is labeled by $12$, the two children will have leaves $n-i$ and $i$
for some $i, 1 \leq i \leq n-1$. This along with the skew rule dictates that
$a_n = \sum_{i=1}^{n-1} b_{n-i} t_i + c_{n-i}t_i+ d_{n-i}t_i$.
Similarly if the root is
$21$ then its left child cannot be $21$ but it can be 12,$41352$ ,$25314$ or a
leaf node. But for trees whose roots are labeled $41352 / 25314$, they can
have any label for any of the five children. Hence we get,$
  a_n  =  t^5_{n - 1} .1 + \Sigma_{i = 2}^{n - 1}  (b_i + c_i +
  d_i) t^5_{n - i},
  b_n = t^5_{n - 1} .1 + \Sigma_{i = 2}^{n - 1}   (a_i + c_i +
  d_i) t^5_{n - i},
  c_n  =  \Sigma_{\left\{ i, j, k, l, m \geq 1| i + j + k + l + m = n
  \right\}} t^5_i t^5_j t^5_k t^5_l t^5_m$ and 
  $d_n  =  \Sigma_{\left\{ i, j, k, l, m \geq 1| i + j + k + l + m = n
  \right\}} t^5_i t^5_j t^5_k t^5_l t^5_m
$
Note that $c_n = d_n$. Since a node labeled $41352 / 25314$ ought to have five
children, $c_{n,} d_n = 0$ for $n < 5$. Summing up $a_n$ and $b_n$ and using
the identity $t^5_i = a_i + b_i + c_i + d_i$ we get the following recurrence for $t^5_n$


\begin{eqnarray*}
    t_n & = & t^5_{n - 1} + \Sigma_{i = 1}^{n - 1} t^5_{n - i} t^5_i + \nonumber \\
    &  & 2 \Sigma_{\left\{ h, i, j, k, l, m \geq 1| h + i + j + k + l + m = n
    \right\}} t^5_h t^5_i t^5_j t^5_k t^5_l t^5_m +  \nonumber \\
    &  & 2 \Sigma_{\left\{ i, j, k, l, m \geq 1| i + j + k + l + m = n
    \right\}} t^5_i t^5_j t^5_k t^5_l t^5_m
   \label{eqn:recurrenceHFO5}
\end{eqnarray*}
Define the ordinary generating function $T (z)$ associated with the sequence $t_n$ to be $T (z) = \Sigma_{n = 1}^{\infty} t_n z^{n - 1}$.
Multiplying the recurrence with $\Sigma_{n = 1}^{\infty} z^{n - 1}$, we get
$T (z) = zT (z) + zT^2 (z) + z^4 T^5 (z) + z^5 T^6 (z) + t_1$.
Substituting $t_1 = 1$, gives the following polynomial equation in $T (z)$,
 $z^5 T^6 (z) + z^4 T^5 (z) + zT^2 (z) + (z - 1) T (z) + 1 = 0$.
Unfortunately this is a polynomial of sixth degree in $T(z)$.
Hence no general solution
is available for its roots, which are needed to obtain the closed form
expression for the above recurrence relation.

Note that in a similar way recurrence relation for any $\tmop{HFO}_k$ can be
constructed. Again it will be a polynomial in $T \left( z \right)$ with degree
$l$ where $l$ is the smallest $l$ such that $\tmop{HFO}_l = \tmop{HFO}_k$.

Even though the above recurrence fails to give a closed form solution
it leads to a natural dynamic programming based algorithm for counting
the number of $\tmop{HFO}_k$ floorplans with $n$ rooms. For example
the recurrence for $\tmop{HFO}_5$ is given by a sixth order recurrence
relation given in Equation~\ref{eqn:recurrenceHFO5}. Hence there is an
$O \left( n^6 \right)$ tabular algorithm computing the value of $t_n$
using dynamic programming which recursively computes all $t^5_i$ for
all $ i < n$ and then computes $t^5_n$ from
Equation~\ref{eqn:recurrenceHFO5}. In general $\tmop{HFO}_k$ has a
recurrence relation of order $k$, and hence the algorithm for $t_n^k$
would run in time $O \left( n^{k + 1} \right)$ using a similar
strategy.

Using the argument which proved existence of an in-decomposable $\tmop{HFO}_k$ floorplan
for any $k$, we can get a simple lower bound on the number of $\tmop{HFO}_k$
floorplans with $n$ rooms which are not $\tmop{HFO}_j$ for any $j < k$. It is
known {\cite{shen2003bounds}} that the number of $\tmop{HFO}_2$ floorplans
with $n$ rooms is $\theta \left( n! \frac{\left( 3 + \sqrt{8} \right)^n}{n^{1.5}}
\right)$. If in the generating tree corresponding to an $\tmop{HFO}_2$
floorplan an in-decomposable $\tmop{HFO}_k$ floorplan is inserted replacing one
of the leaves (to be uniform, say the right most leaf), the resulting
generating tree would be of order $k$ and hence by Theorem \ref{thm:skwdgt},
would correspond to an $\tmop{HFO}_k$ floorplan. Hence the number of
$\tmop{HFO}_k$ floorplans with $n$ rooms which are not $\tmop{HFO}_l$ is at
least the number of generating trees of order $2$ with $n - k + 1$ leaves. And
the number of generating trees of order $2$ with $n$ leaves equals the number
of $\tmop{HFO}_2$ floorplans with $n$ rooms thus giving the following
exponential lower bound.

\begin{observation}
  For any $k \geq 7$, the number of $\tmop{HFO}_k$ floorplans \ with $n$ rooms which are not
  $\tmop{HFO}_j$ for any $j < k$ is at least
  \[ \frac{(n-k)!\left( 3 + \sqrt{8} \right)^{n - k}}{\left( n - k \right)^{1.5}}
  \]
  
\end{observation}

\section{Algorithm for membership}
\label{sec:algorithmForMembership}

For arriving at an algorithm for membership in $\tmop{HFO}_k$ we note that
if a given permutation is Baxter then it is $\tmop{HFO}_k$ for some $k$.
And if it is $\tmop{HFO}_k$ by Theorem \ref{thm:skwdgt} there exits an order
$k$ generating tree corresponding to the permutation. By Theorem~\ref{thm:skwdgt}
the generating tree also captures the block decomposition of
the permutation. For sake of brevity we defer the formal description of 
the algorithm to the Appendix (see Algorithm~\ref{alg:algor-recogn}).
 Our algorithm identifies the block-decomposition
corresponding to the generating tree of order $k$, level by level. It can be
thought of as a \tmtextit{deflating} algorithm, i.e., it finds the block
decomposition which when \tmtextit{inflated} gives the input permutation. The
algorithm first identifies the \tmtextit{blocks} length at most $k$ in the
input permutation which corresponds to the leaves of the generating tree. Upon
finding a \tmtextit{block} algorithm replaces the \tmtextit{block} with the interval
$\left[ i, j \right]$ where $\left[ i, j \right]$ are the elements of the
\tmtextit{block}. Hence after the first round the input permutation is changed
to an ordered arrangement of entries which are intervals $\left[ i, j \right]$ for some $i \leq j$. And in the
subsequent round the algorithm tries to identify the \tmtextit{blocks} of at
most $k$ such entries. The rounds continue until the permutation is reduced
to a single entry $\left[ 1, n \right]$ or till a round fails to identify a
\tmtextit{block} of length at most $k$. If the given permutation is reduced to
a single permutation at the end, the algorithm guarantees that there is a
\tmtextit{block} decomposition of the given permutation where the maximum
in-decomposable \tmtextit{block} is of length $k$. Hence if the permutation,
after running the algorithm is reduced to a single permutation, it is indeed
$\tmop{HFO}_k$. And if the permutation is $\tmop{HFO}_k$ then there is a
generating tree of order $k$ corresponding to it, and this guarantees that the
algorithm would be able to reduce it to a single permutation by the level by
level compression strategy\footnote{Figure~\ref{fig:algorun} given in Appendix illustrates the decompositions identified
at each round of the algorithm on input $13274685$ checking whether it is
$\tmop{HFO}_5$ or not.}.

Note that checking if a set $S$ of $k$ elements form a range can be checked in
constant time for a fixed value of $k$ by subtracting from each element
$\min_{i \in S} \left( i \right) - 1$ and then checking if the elements follow
any of the $k$! arrangements. We can also check if a set of $k$ elements form
a Baxter permutation for a fixed $k$ in constant time by checking if their
rank ordering is equivalent to any one of the Baxter permutations of length
$k$(whose number is bounded by number of permutations, $k$!). After each round
of algorithm at least one non-trivial block-decomposition is identified and
\tmtextit{deflated}. Hence in each round the number of nodes in the
corresponding generating tree reduces by at least one. Note that if the input
permutation is not $\tmop{HFO}_k$, then algorithm progresses only till it can
find a block-decomposition which can be \tmtextit{deflated}. Hence the number of rounds
is linear in the number of nodes of the generating tree. And each round
takes at most linear time.
Since any tree
with $n$ leaves where each internal node has degree at least $2$ has, at most
$n - 1$ internal nodes, the total running time is $cn(2n-1)$.
Hence the above algorithm runs in $O(n^2)$ time for a predetermined value of
$k$.

For a fixed $k$ we can also achieve linear time for membership owing to a new
fixed parameter algorithm of Marx and Guillemot~\cite{marx_fpt_pp} which given
two permutations $\sigma \in S_k$ and $\pi \in S_n$ checks if $\sigma$ avoids
$\pi$ in time $2^{O(k^2 \log k)}n$ and a linear time algorithm for recognizing 
Baxter permutations by Hart and Johnson~\cite{linearbaxter}.
Both results~(\cite{marx_fpt_pp,linearbaxter}) are highly non-trivial and deep.
 Theorem~\ref{thm:hfokcharacterization} 
guarantees that it is enough to ensure that $\pi$ is Baxter and $\pi$ and avoids
simple permutations of length $k+1$ and exceptionally simple permutations of length $k+2$.
Using the algorithm given by Hart and Johnson~\cite{linearbaxter} we can check 
in linear time whether a given permutation is Baxter or not.
Since there are at most $(k+1)!$ simple permutations of length $k$ and at most $(k+2)!$
exceptionally simple permutations of length $(k+2)$ using the algorithm given in 
\cite{marx_fpt_pp} as a sub-routine we can do the latter in $O((k+2)!2^{c (k+2)^2 \log (k+2)} n)$ time.
 Since $k$ is a fixed constant we get a linear time algorithm.

If the value $k$ is unknown Algorithm~\ref{alg:algor-recogn} can be used to get an $O(n^4)$
algorithm with a few modifications to find out the minimum $k$ for which the input permutation is
$\tmop{HFO}_k$. The first
modification is to make the algorithm check if the input permutation $\pi$ is
Baxter permutation. If it is not, it cannot be $\tmop{HFO}_j$ for any $j$  and hence
is rejected. If it is a Baxter permutation then it is $\tmop{HFO}_k$ for some
$k \leq n$. And in each round we check for the minimum $j, 1 < j \leq |S|$ for which
the top $j$ elements form a range $[l,m]$ and is a Baxter permutation shifted by $l$.
Checking if a
permutation is Baxter takes $O (n^2)$ time. And as earlier there are at most $2n$ rounds.
In each round checking whether a set of elements forms a range takes $O(n \log_2 n)$ time
and checking if the resulting permutation is Baxter takes $O(n^2)$ time. Since 
there are at most $n$ elements in the stack at any time, the  worst case cost of
a round is $O(n^3)$. Hence the running time of the algorithm is $O(n^4)$.

We also note that algorithm for membership becomes much simpler if you want to
check whether a permutation is $\tmop{HFO}_k$ for a fixed $k$. Because of Theorem~\ref{thm:skwdgt}
for any $\tmop{HFO}_k$ permutation $\pi$ there is a unique \textsgto{k}, $T_\pi$ such that the tree yields block decomposition of $\pi$
when thought 
of as a parse tree. It is easy to see that the recurrence for generating trees 
or order $k$ based on what root is labeled by gives a context free grammar
for generating such tree. See Appendix~\ref{subsec:ContextFreeGrammarApproach} 
for the details of the algorithm based on the context free grammar approach.

\section{Closure properties of Baxter permutations}\label{sec:closurehfok}

Only recently it has been proved that Baxter permutations are closed under inverse~\cite{law2011hopf}. The proof in~\cite{law2011hopf} uses an argument based on permutations and patterns.
We give a simple alternate proof of this fact using the
geometrical intuition derived from mosaic floorplans. We prove that the floorplan
obtained by taking a mirror image of a floorplan along the horizontal axis
is a floorplan whose permutation (under the bijection of Ackerman) is the
inverse of the permutation corresponding to the starting floorplan.

    The intuition is that  when the floorplan's mirror image
    about the horizontal axis is taken, it
    does not change the relationship between  
    two rooms if one is to the left of the other. But if a room is below the other, 
    it flips the relationship between the corresponding rooms.
    For any Baxter permutation $\pi$ and two indices $i,j$ where $i < j$, 
    if $\pi[i] < \pi[j]$, since $\pi[i]$ appears before $\pi[j]$
    by the property of the algorithm FP2BP  $\pi[i]$ is to the left of $\pi[j]$
    in $\pi_{f}$.
    In the inverse of $\pi$, $\pi^{-1}$ indices $\pi[i]$ and $\pi[j]$ will
    be mapped to $i$ and $j$ respectively. Hence if $\pi^{-1}$ is Baxter, or equivalently there is a mosaic floorplan
    corresponding to $\pi^{-1}$, $\pi^{-1}_{f}$, the rooms
    labeled by $i$ and $j$ will be such that $i$ precedes $j$ 
    in the top-left deletion ordering(as $i<j$) and also in bottom
    left deletion ordering(as $\pi[i]<\pi[j]$). Hence $i$ is to the left of $j$
    in $\pi^{-1}_{f}$. If $\pi[i] > \pi[j]$, since $\pi[i]$ appears before $\pi[j]$ by
    by the property of the algorithm FP2BP, $\pi[i]$ is below $\pi[j]$
    in $\pi_{f}$.
    In the inverse of $\pi$, $\pi^{-1}$ indices $\pi[i]$ and $\pi[j]$ will again
    be mapped to $i$ and $j$ respectively. Hence if there is a mosaic floorplan
    corresponding to $\pi^{-1}$, $\pi^{-1}_{f}$, the 
    rooms labeled by $i$ and $j$ will be such that $i$ precedes $j$ 
    in the top-left deletion ordering(as $i<j$) but in bottom
    left deletion ordering $j$ precedes $i$(as $\pi[i]<\pi[j]$). Hence $i$ is above $j$
    in $\pi^{-1}_{f}$. Thus mirror image about horizontal axis satisfies all 
    these constraints on the rooms.
    
    For the formal proof of closure under inverse we use the following three lemmas. 
    For sake of brevity we defer the proofs to the Appendix (see~\ref{app-lemma:commFlipDeletion},
    \ref{app-lemma:iextractilabel} and~\ref{app-lemma:ilabeliextract}).

\begin{lemma}\label{lemma:commFlipDeletion}
For any mosaic floorplan $f$, the floorplan obtained by deleting a room
from the bottom-left corner 
of $f$ and then taking a mirror image about horizontal axis is \textit{equivalent}
to the floorplan obtained by taking a mirror image of $f$ about horizontal
axis and then deleting a room from the top-left corner.
\end{lemma}

\begin{lemma}\label{lemma:iextractilabel}
For any mosaic floorplan $f$, let $g$ be the floorplan obtained from $f$ by taking
a mirror image about the horizontal axis. Then the $i$th ($1\leq i \leq n$) room
deleted from $f$ during the extraction phase of algorithm $FP2BP$ on $f$ is the $i$th room to
be deleted in the labelling phase of algorithm $FP2BP$ on $g$.
\end{lemma}

\begin{lemma}\label{lemma:ilabeliextract}
For any mosaic floorplan $f$, let $g$ be the floorplan obtained from $f$ by taking
a mirror image about the horizontal axis. Then the $i$th ($1\leq i \leq n$) room
deleted from $f$ during the labeling phase of algorithm $FP2BP$ on $f$ is the $i$th room to
be deleted in the extraction phase of algorithm $FP2BP$ on $g$.
\end{lemma}

Now we can proceed to the proof of closure of Baxter permutations under inverse.
\begin{theorem}\label{thm:closureunderinverse}
  Given a mosaic floorplan $f$, the floorplan $g$ obtained by taking the
  mirror image of $f$ about the horizontal axis is such that $\pi_f^{-1} =
  \pi_g$ where $\pi_f, \pi_g$ are the Baxter permutations corresponding to the
  mosaic floorplans $f$ and $g$ respectively.
\end{theorem}

\begin{proof}
Once again note that taking the mirror image of a mosaic floorplan results in a mosaic
floorplan(as cross junctions do not appear through a rotation). 
From the definition algorithm $FP2BP$ for any $i$, $\pi_f[i]=j$ is the $i$th room to
deleted in the extraction phase of $FP2BP$ on $f$. And $\pi_g^{-1}(i)$ is the $i$th room
to be deleted from $g$ in the labeling phase of $FP2BP$. By Lemma~\ref{lemma:iextractilabel}
these rooms are one and the same. By Lemma~\ref{lemma:ilabeliextract}, $j$th 
room to be labeled in $f$ is same as the $j$th room to be extracted in $g$. That is $\pi_g[j]=i$, 
which means that $\pi_g^{-1}(i)=j=\pi_f[i]$ for any $i$. Hence $\pi_g$ is the inverse of $\pi_f$.

   Figure \ref{fig:GeomEquivInverse} illustrates the above mentioned link
   between inverse and the geometry.
   
   \begin{figure}
   \centering 
   \includegraphics[scale=0.25]{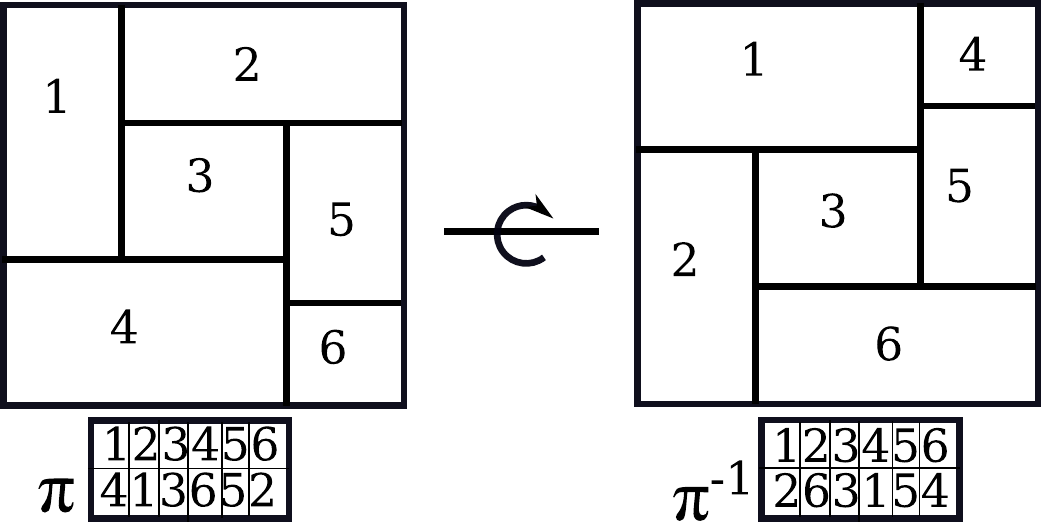}

   \caption{Obtaining a mosaic floorplan corresponding to the inverse of a Baxter permutation}
   \label{fig:GeomEquivInverse}
   \end{figure}

\end{proof}

From Theorem~\ref{thm:closureunderinverse} we get the following corollary,

\begin{corollary}
For any $k\in \mathbb{N}$, if $\pi \in \tmop{HFO}_k$ then so is $\pi^{-1}$
\end{corollary}
\begin{proof}
Since $\pi \in \tmop{HFO}_k$, $\pi$ is also a Baxter permutation. And according
to Theorem~\ref{thm:closureunderinverse} $\pi^{-1}$ is also a Baxter permutation
whose corresponding floorplan (under the bijection of Ackerman et. al) is 
 obtained by taking the  mirror image of $f_{\pi}$ about the horizontal axis.
Theorem~\ref{thm:generating_tree_correspondence} guarantees that there is a 
generating tree $T_{\pi}$ of order $k$ corresponding to $f_{\pi}$. The nodes of the tree
are labeled by Baxter permutations of length at most $k$. Now obtain a new tree $T'$
by relabeling each node, starting from root, by inverse of the permutation labeling the node and moving
the children of the node accordingly.
For a node $u \in T_{\pi}$ the corresponding node $u' \in T'$ gets labeled by
inverse of the Baxter permutation labeling $u$. Note that this still is Baxter 
permutation of length at most $k$. Hence the generating tree $T'$ represents 
an $\tmop{HFO}_k$ permutation because of Theorem~\ref{thm:generating_tree_correspondence}.
It is not hard to verify that the floorplan represented by $T'$ is the mirror image
of $f_{\pi}$ taken about the horizontal axis.

\end{proof}

We also observe that there is a geometric interpretation for reverse of a Baxter permutation. Note that it is easy to see that
Baxter permutations are closed under reverse because the patterns they avoid are reverses
of each other ($3142/2413$). We observe, without giving a proof,
that for a Baxter permutation $\pi$ its reverse
$\pi^{r}$ corresponds to the mosaic floorplan that is obtained by first rotating by $90^\circ$ clockwise and then by taking a mirror
image along the horizontal axis. See
Figure~\ref{fig:geomequivrev} in the Appendix for an illustration of this link.



\section{Open Problems}\label{sec:openproblems}

One natural open problem arising from this work is that of exact formulae for the
number of $\tmop{HFO}_k$ floorplans. The only $k$ for which exact count is known is $k=2$. Our proof of closure under inverse for Baxter permutation gives rise to the following open problem. For a class of permutations characterized by pattern avoidance, like Baxter permutations, to be closed under inverse is it enough that the forbidden set of permutations defining the class is closed under
inverse.

\bibliography{references}
\bibliographystyle{plain}

\appendix
\section{Appendix}

\subsection{Example Floorplans}
\label{app:figures}
\begin{figure}[H]
\centering
\includegraphics[scale=0.2]{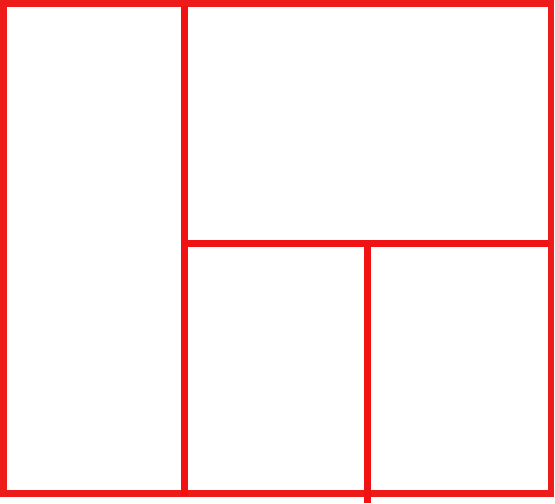}
\caption{A slicing($\tmop{HFO}_2$) floorplan with $4$ rooms}
\label{fig:slicingmoretwo}
\end{figure}

\begin{figure}[H]
\centering
\includegraphics[scale=0.2]{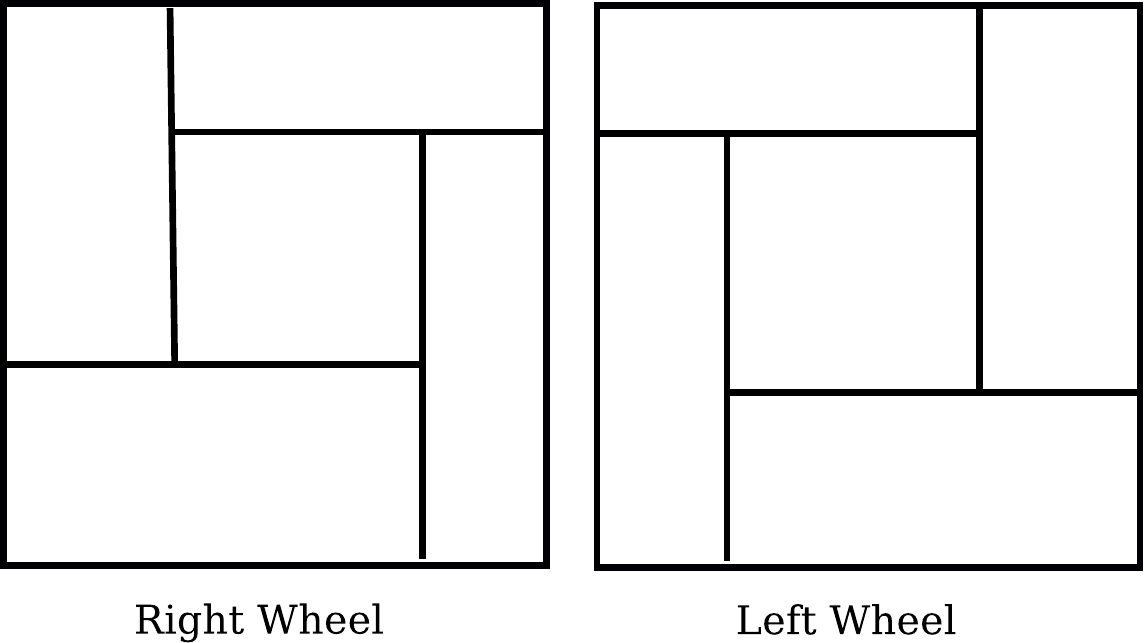}
\caption{Only non $\tmop{HFO}_2$ floorplans with at most five rooms}
\label{fig:wheels}
\end{figure}

\begin{figure}[H]
\centering
\includegraphics[scale=0.2]{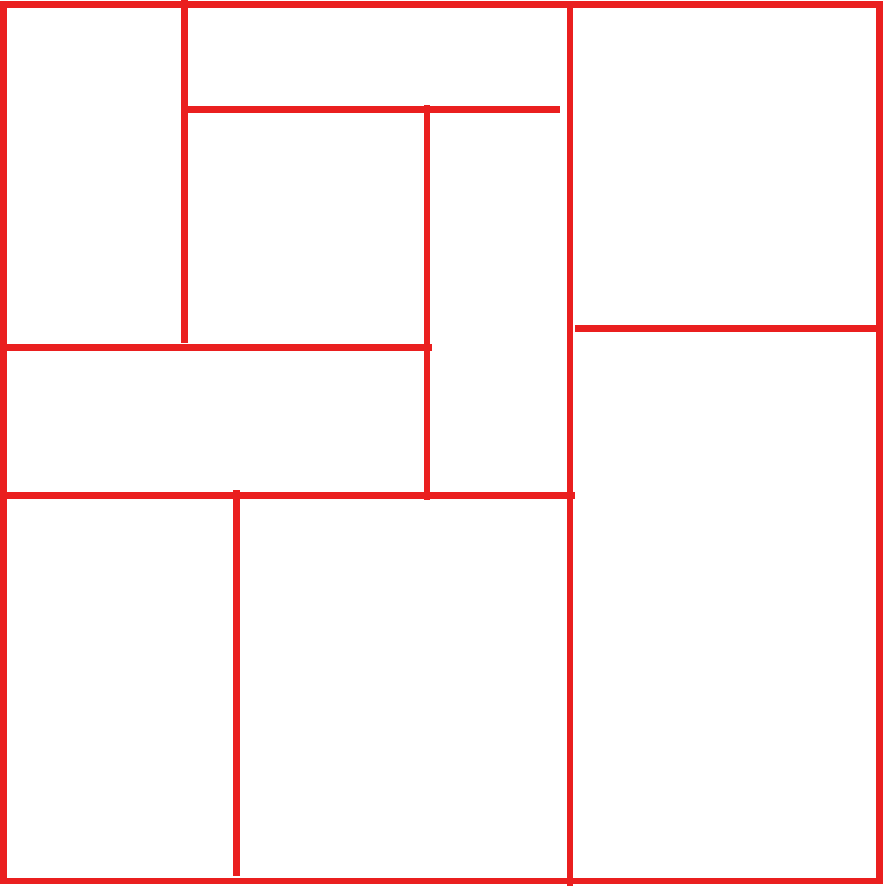}
\caption{An $\tmop{HFO}_5$ which is not $\tmop{HFO}_2$}
\label{fig:hfofiveninerooms}
\end{figure}

\begin{figure}[H]
\centering
\includegraphics[scale=0.2]{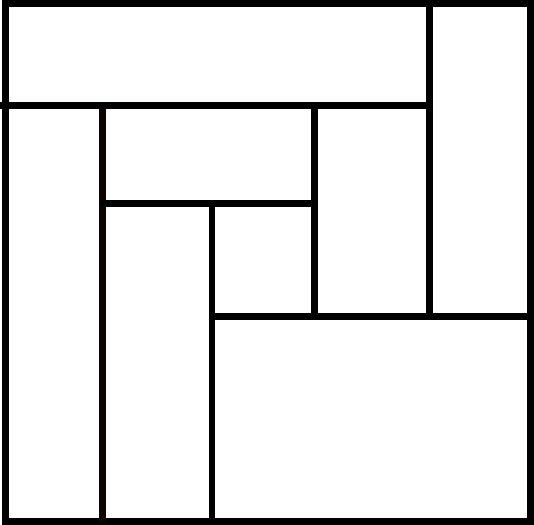}
\caption{An $\tmop{HFO}_8$ which is not $\tmop{HFO}_7$}
\label{fig:indhfo8}
\end{figure}

\begin{figure}[H]
\centering
\includegraphics[scale=0.2]{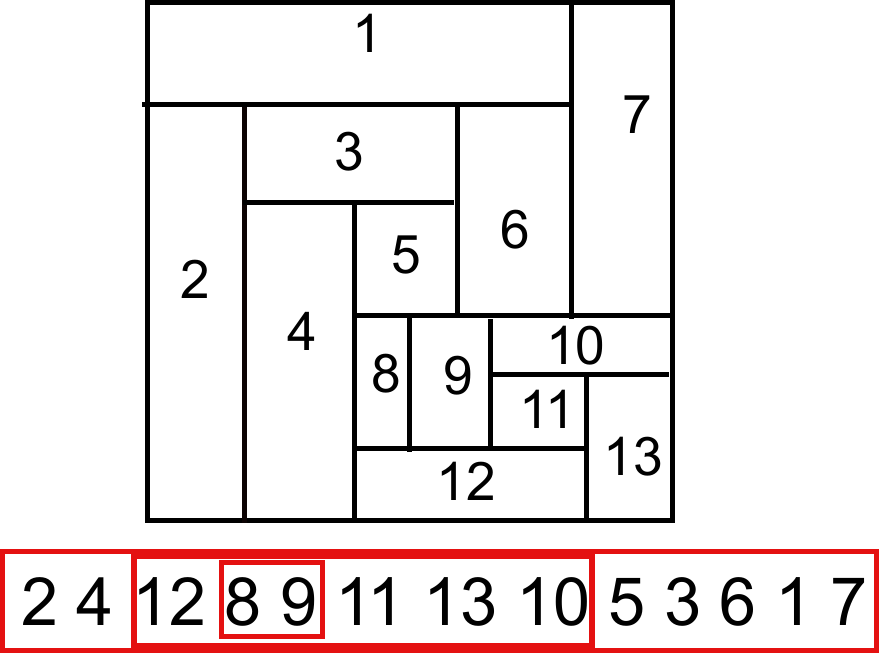}
\caption{An $\tmop{HFO}_8$ with $10$ rooms, with the blocks identified}
\label{fig:hfo813}
\end{figure}

We provide example floorplans for various $\tmop{HFO}_k$. Figure~\ref{fig:slicingmoretwo}
shows an $\tmop{HFO}_2$ with more than $2$ rooms. Figure~\ref{fig:wheels} shows the smallest
non-slicing(non $\tmop{HFO}_2$) floorplan (and it contains five rooms). The structure is called
a ``pin-wheel'' and there are only two of them, one right rotating and another left rotating
as shown in Figure~\ref{fig:wheels}. Figure~\ref{fig:hfofiveninerooms} shows an $\tmop{HFO}_5$ which
is not $\tmop{HFO}_2$ by slicing a wheel. Figure~\ref{fig:indhfo8} shows an $\tmop{HFO}_8$ 
which is not an $\tmop{HFO}_7$ floorplan. Figure~\ref{fig:hfo813} shows another $\tmop{HFO}_8$
with $10$ rooms.

\clearpage

\subsection{Proof's omitted from the main paper}

\begin{corollary}
  A mosaic floorplan $f$ is in-decomposable if and only if the Baxter
  permutation $\pi_f$ corresponding to it is block
  in-decomposable.\label{app-coro:mosaicindecomp}
\end{corollary}

\begin{proof}
  If $f$ is decomposable then Lemma \ref{lemma:FPMosaicInjection} guarantees
  that $\pi_f$ is also decomposable. Suppose $\pi_f$ is decomposable say
  $\pi_f = \rho \left( \sigma_1, \ldots, \sigma_k \right)$, then Lemma $1$
  guarantees that the floorplan $f'$ obtained by insertion of $f_{\rho}$ by
  $f_{\sigma_1}, \ldots, f_{\sigma_k}$ (the floorplans corresponding to
  permutations $\rho, \sigma_1, \ldots, \sigma_k$) also corresponds to
  $\pi_f$. Since mosaic floorplans are in bijective correspondence with Baxter
  permutations it must be the case that $f \equiv f'$.
\end{proof}

\begin{lemma}
  If $\pi = \sigma \left( \rho_1, \ldots, \rho_k \right)$, then $\pi$ contains
  all patterns which any of $\sigma, \rho_1, \rho_2, \ldots, \rho_k$
  contains.\label{app-lemma:blockdecomposition} 
\end{lemma}

\begin{proof}
  Let $\pi = \sigma \left( \rho_1, \ldots, \rho_k \right)$. \ If $\sigma$ has
  a \tmtextit{text} indexed by $i_1, \ldots, i_m$ matching a pattern $\alpha
  \in S_m$, then by taking an arbitrary element from each block corresponding to $\rho_{i_1},
  \ldots, \rho_{i_m}$ in $\pi$, a \tmtextit{text} matching
  $\alpha$ is obtained. This is because \tmtextit{inflation} orders
  blocks corresponding to
  $\rho_1, \rho_2, \ldots, \rho_k$ by $\sigma$. 
  Similarly if some
  $\rho_i$ has a \tmtextit{text} matching some pattern $\alpha$ then $\pi$
  having the \tmtextit{block} corresponding to $\rho_i$  contains the same \tmtextit{text}
  as this block preserves the
  relative ordering of elements inside the \tmtextit{block} according to
  $\rho_i$.
\end{proof}

\begin{lemma}
  If $\pi = \sigma \left( \rho_1, \ldots, \rho_k \right)$, then any 
  block in-decomposable pattern in $\pi$ has a matching text which is completely 
  contained in one of $\sigma,\rho_1,\rho_2,\ldots,\rho_k$.
  \label{app-lemma:indecomposable_implies_inflation}
\end{lemma}

\begin{proof}
  Recall that a pattern is block in-decomposable if it does not have any non-trivial blocks.
  Suppose if the pattern $\gamma$ which is block in-decomposable has at least two matching
  text elements from any $\rho_i$ and also contains a matching text element from a $\rho_j$
  for $j \neq i$, then the matching text elements from $\rho_i$ forms a non-trivial block of 
  $\gamma$ as $\rho_i$'s form blocks in $\pi$ by definition of inflation. Hence $\gamma$ can
  have either at most one matching text element from each $\rho_i$ in which case $\sigma$
  also contains the text by definition of inflation, or it is completely contained in one 
  of $\rho_i$'s. Hence the theorem.
\end{proof}

\begin{lemma}\label{app-lemma:ilabeliextract}
For any mosaic floorplan $f$, let $g$ be the floorplan obtained from $f$ by taking
a mirror image about the horizontal axis. Then the $i$th ($1\leq i \leq n$) room
deleted from $f$ during the labeling phase of algorithm $FP2BP$ on $f$ is the $i$th room to
be deleted in the extraction phase of algorithm $FP2BP$ on $g$.
\end{lemma}

The proof is similar to the proof of earlier lemma.

\begin{theorem}
  For any $k \geq 7$, there exists a floorplan $f$ which is in $\tmop{HFO}_{k
  + 2}$ but is not in $\tmop{HFO}_{l}$ for any $l \leq k+1$
  \label{app-thm:hfokseparataion}
\end{theorem}

\begin{proof}
  We generalize the above procedure to one which obtains $\pi_{k + 2}$ from a
  $\pi_k$ which is an in-decomposable $\tmop{HFO}_k$ having $\pi \left[ n - 1
  \right] = n$ and $\pi \left[ n \right] = 2$, as follows $\pi_{k + 2} \left[
  1 \right] = n + 1$, $\pi_{k + 2} \left[ i \right] = \pi_k \left[ i - 1
  \right], 2 \leqslant i \leqslant n$, $\pi_{k + 2} \left[ n + 1 \right] = n +
  2$ and $\pi_{k + 2} \left[ n + 2 \right] = 2$. If $\pi_k$ is a Baxter
  permutation which is also \tmtextit{simple} then $\pi_{k + 2}$ is also
  Baxter and \tmtextit{simple}. Before discussing the proof we can see why
  this help prove the theorem. For $k = 5$, $\pi_5=41352$ satisfies
  the conditions, and the procedure makes sure that successive $\pi_k$'s
  obtained also have $\pi \left[ n - 1 \right] = n$ and $\pi \left[ n \right] =
  2$ allowing us to separate any odd $k$ and $k + 2$. For even $k$ we observe
  that $\pi_8 = 7 5 1 4 6 3 8 2$ satisfies the necessary
  conditions hence proving a similar theorem for even number $k, k \geqslant 8$.
  
  We prove that $\pi_{k + 2}$ obtained as above is both \tmtextit{simple} and
  Baxter. Since $\pi_k$ is simple, if there is a nontrivial \tmtextit{block} in $\pi_{k+2}$
  it must include one of the newly inserted elements (i.e., $n+1$ or $n+2$) or the moved element (i.e., $2$). 
  But if the non-trivial \tmtextit{block} includes $n + 2$, it must
  involve $n + 1$ as there are no elements greater than $n + 2$, but in this
  case the block will have to include $1$ as it is between $n + 2$ and $n + 1$,
  but if it includes $1$ it becomes a \tmtextit{trivial block} which is the
  entire permutation. Similarly if the block involves $n + 1$ it must include
  $n$, or $n + 2$ but in both cases $1$ must be included, thus making the
  block trivial. If it involves $2$ it must also include $n + 2$ as it is the
  only number adjacent to it, but we have already proved that the
  non-\tmtextit{trivial block} cannot contain $n + 2$. Hence $\pi_{k + 2} $ is
  simple.
  
  To see why $\pi_{k + 2}$ is Baxter, assume to the contrary there exists a
  \tmtextit{text} at indices $i_1, i_2, i_3, i_4$ which \tmtextit{matches}
  $3142 / 2413$ with $\left| \pi \left[ i_1 \right] - \pi \left[ i_4 \right]
  \right| = 1$, then at least one of $\pi \left[ i_1 \right], \pi \left[ i_2 \right],
  \pi \left[ i_3 \right], \pi \left[ i_4 \right]$ must be from $\left\{ n + 1,
  n + 2 \right\}$. Because otherwise the same \tmtextit{text} would appear in
  $\pi_k$ also as we have not changed the relative positions of elements in
  $\pi_k$ while constructing $\pi_{k + 2}$. If it involves $n + 2$ it must be
  the case that it matches $4$ as it is the largest element in $\pi_{k + 2}$.
  Since there are no two elements after $n + 2$ in one-line form of $\pi_{k+2}$
  it cannot match $2413$, and if
  it matches $3142$ it must be the case that $2$ in $\pi_{k + 2}$ matches the
  $2$ in the \tmtextit{pattern} as it is the only number to the right of $n +
  2$ in the one-line form of $\pi_{k+2}$. 
  But if $2$ matches $2$ in the pattern, 1 of $\pi_{k + 2}$ must match $1$
  of the \tmtextit{pattern} and hence 3 of $\pi_{k + 2}$ must match $3$ in the
  \tmtextit{pattern} as we need absolute difference of numbers matching $3$
  and $2$ to be $1$. But in this case we can replace $n + 2$ in the
  \tmtextit{text} with $n$ to get a new \tmtextit{text} which is also present in
  $\pi_k$ matching $3142$, contradicting our assumption that $\pi_k$ is
  Baxter. Hence it cannot involve $n + 2$, but if it involves $n + 1$ by
  virtue of it having only element greater than itself in $[n+2]$, it cannot match $1$ or $2$ in
  the \tmtextit{pattern}. 
  Also it cannot match $4$ in $3142/2413$ as there is no number to the left
  of $n+1$ in one-line form of $\pi_{k+2}$.
  Similarly it cannot match $3$ in $2413$ as there is nothing to the left of $n+1$. Thus it can only match $3$ in $3142$.
  But if $n + 1$ matches $3$ then $n+2$ must match $4$ as it is 
  the only number greater than $n+1$ and thus forcing $n$ to match $2$ as the absolute value of difference is $1$.  
  But $n+2$ matching $4$ lies to the right of $n$ matching $2$, rendering such a case impossible.
  
  Thus the permutation obtained by above construction is both simple and Baxter. Hence it contains a pattern which is a simple permutation of length $k+2$ (the whole permutation). This fact along with Theorem~\ref{thm:hfokcharacterization} implies that the permutation thus obtained cannot be $\tmop{HFO}_\ell$ for $\ell < k+2$.
\end{proof}

\begin{lemma}\label{app-lemma:commFlipDeletion}
For any mosaic floorplan $f$, the floorplan obtained by deleting a room
from the bottom-left corner 
of $f$ and then taking a mirror image about horizontal axis is \textit{equivalent}
to the floorplan obtained by taking a mirror image of $f$ about horizontal
axis and then deleting a room from the top-left corner.
\end{lemma}

\begin{proof}
Ackerman et al. proved that deletion operation does not change the relation
between any two blocks (see proof of, \cite[Observation 3.4]{Ackerman20061674})
of a mosaic floorplan. And the top-left room of the
floorplan obtained by taking a mirror image of $f$ about horizontal axis is 
the bottom-left room of of $f$. This combined with the fact that 
image about the horizontal axis does not change the relationship between  
two rooms if one is to the left of other, but flips relationship between two
rooms if one is below the other proves the theorem.
\end{proof}

\begin{lemma}\label{app-lemma:iextractilabel}
For any mosaic floorplan $f$, let $g$ be the floorplan obtained from $f$ by taking
a mirror image about the horizontal axis. Then the $i$th ($1\leq i \leq n$) room
deleted from $f$ during the extraction phase of algorithm $FP2BP$ on $f$ is the $i$th room to
be deleted in the labelling phase of algorithm $FP2BP$ on $g$.
\end{lemma}

\begin{proof}
    Note that $g$ is also a mosaic floorplan as reflections cannot introduce 
    cross junctions.
    The proof is an induction on $i$.
    When $i=1$ the first room to be deleted from $f$ in 
    the extraction phase is the bottom-left room of $f$. Clearly it is the top-left room
    in the floorplan $g$.
    In the labelling phase of $FP2BP$ on $g$ the first room to be
    labelled is the top-left room. Hence they are one and the same. 
    
    Assume that the hypothesis
    is true for $i$, that is the $i$th room to be deleted from $f$ is the $i$th room labelled
    in $g$. Let $f'$ be the floorplan obtained from $f$ by deleting
    $i$ rooms from the bottom-left corner. Similarly $g'$ is obtained from $g$ by deleting
    $i$ rooms from the top-left corner. Repeated application of Lemma~\ref{lemma:commFlipDeletion}
    implies that mirror image $f'$  about horizontal axis is \textit{equivalent} to $g'$. 
    The bottom-left room of $f'$ is the $i+1$th
    room to be deleted from in $f$ in the extraction phase of $FP2BP$. 
    The bottom-left room of $f'$ is the top-left room of $g'$ as mirror image
    of $f'$ about horizontal axis is \textit{equivalent} to $g'$. Hence 
    the it is the room to be deleted from the top-left corner of $g'$. But 
    by definition of $g'$ it is the $i+1$th room to be labeled in the labeling phase 
    of $FP2BP$ on $g$. Hence the theorem.
\end{proof}

\subsection{Pseudo-code for the Algorithm for membership}
\label{subsec:pseudoCodeAlgoMembership}
\begin{algorithm}[H]
 \SetKwData{Stck}{S}
 \SetKwData{Stckk}{SS}
 \SetKwData{Deflated}{Deflated}
 \SetKwData{Rng}{R}
 \SetKwData{Top}{top}
 \SetKwData{I}{i}
 \SetKwData{J}{j}
 \SetKwData{K}{k}
 \SetKwData{Lvar}{l}
 \SetKwData{Mvar}{m}
 \SetKwData{N}{n}

 \SetKw{KwDownTo}{downto}
 \SetKw{KwAccept}{Accept}
 \SetKw{KwReject}{Reject}

 \SetKwFunction{Push}{push}
 \SetKwFunction{Pop}{pop}
 \SetKwFunction{Range}{Range}
 \SetKwFunction{Size}{size}

 \SetKwInOut{Input}{Input}

 \Input{A permutation $\pi$ of length $n$}
 \BlankLine
 Stack \Stck$\leftarrow \phi$\;
 Stack \Stckk$\leftarrow \phi$\;
 Boolean \Deflated$\leftarrow$ \textbf{true}\;

 \For{\I$=1$ \KwTo \N} {
    \Stck.\Push{$[\pi[i],\pi[i]]$}\;
 }
 \While{ \Deflated \textbf{AND} \Stck.\Size{} $\ne 1$}{
  \Deflated = \textbf{false}\;
    \While{There exists a \Lvar,\J, such that \J is the maximum such number less
    than $k$ for  which top \J$+1$ elements of \Stck is $[$\Lvar,\Lvar$+$\J$]$} {
     \If{\Stck$[\Top \dots(\Top-\J)]$ is a Baxter permutation shifted by $\Lvar-1$}{
          \Rng = $[$\Lvar,\Lvar$+$\J$]$\;
          \If{\Lvar $\ne $ \Lvar$+$\J }{
          \Deflated = \textbf{true}\;
            }
          \For{\Mvar$=$\J \KwDownTo $0$ } {
             \Stck.\Pop{}\;
          }
          \Stckk.\Push{\Rng}\;
     }
    }
    \While{\Stckk.\Size{} $\ne 0$} {
      \Stck.\Push{\Stckk.\Pop{}}\;
    }
 }
 \eIf{\Stck.\Size{}$=1$} {
 \KwAccept\;
 } {
  \KwReject\;
 }

 \caption{Algorithm for checking if a permutation is $\textrm{HFO}_k$}
 \label{alg:algor-recogn}
\end{algorithm}

The following figure demonstrates the identification of blocks as employed
by the above algorithm on an $\tmop{HFO}_5$ permutation.

\begin{figure}[H]

\centering

\includegraphics[scale=0.2]{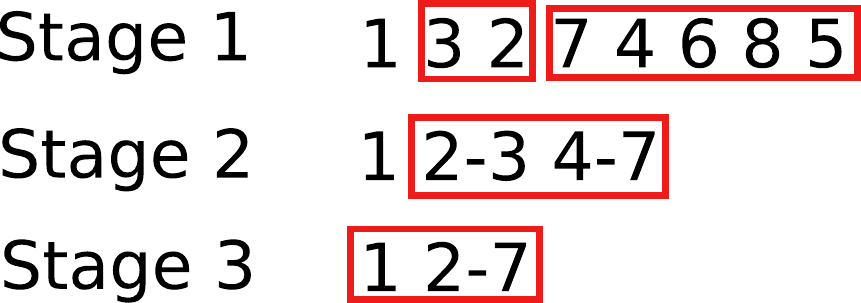}

\caption{Running of algorithm for $\tmop{HFO}_5$ recognition}
\label{fig:algorun}

\end{figure}

\clearpage

\subsection{Context Free Grammar Approach for checking membership}
\label{subsec:ContextFreeGrammarApproach}

The context free grammar 
corresponding to \textsgto{5}, which yields block-decomposition of all
$\tmop{HFO}_5$ permutations is given below.


\begin{grammar}
    <tree-node> ::= <vertical-slice> 
                \alt <horizontal-slice>
                \alt <right-rotating-wheel>
                \alt <left-rotating-wheel>
                \alt `1'
    
    <vertical-slice> ::= `12 [' <left-skew> <tree-node> `]'
    
    <left-skew> ::= <horizontal-slice>
                \alt <right-rotating-wheel>
                \alt <left-rotating-wheel>
                \alt `1'
                
    <horizontal-slice> ::= `21 [' <right-skew> <tree-node> `]'
                      
    <right-skew> ::= <vertical-slice>
                 \alt <right-rotating-wheel>
                 \alt <left-rotating-wheel>
                 \alt `1'
                 
    <left-rotating-wheel> ::= `25314 [' <tree-node> <tree-node> <tree-node> <tree-node> <tree-node> `]'
    
    <right-rotating-wheel> ::= `41352 [' <tree-node> <tree-node> <tree-node> <tree-node> <tree-node> `]'
\end{grammar}

Now by using CYK-algorithm~\cite{Younger} one can check whether a block-decomposition
is generated by a \textsgto{5}, in time $O(n^3)$. And to produce the block-decomposition
of the given permutation a similar stack based algorithm can be used. The algorithm will run in time order of
number of blocks in the decomposition, which is at most $2n$, as the number of nodes in the 
generating tree and the number of blocks in the corresponding Baxter permutation are the same.
Hence for a fixed $k$ this approach of finding the block-decomposition and using CYK-algorithm
 to see if the corresponding block decomposition is generated, 
takes $O(n^3)$ time.

\subsection{The equivalence between reverse of a Baxter permutation and rotation on a mosaic floorplan}
\begin{figure}{H}
   \centering 
   \includegraphics[scale=0.2]{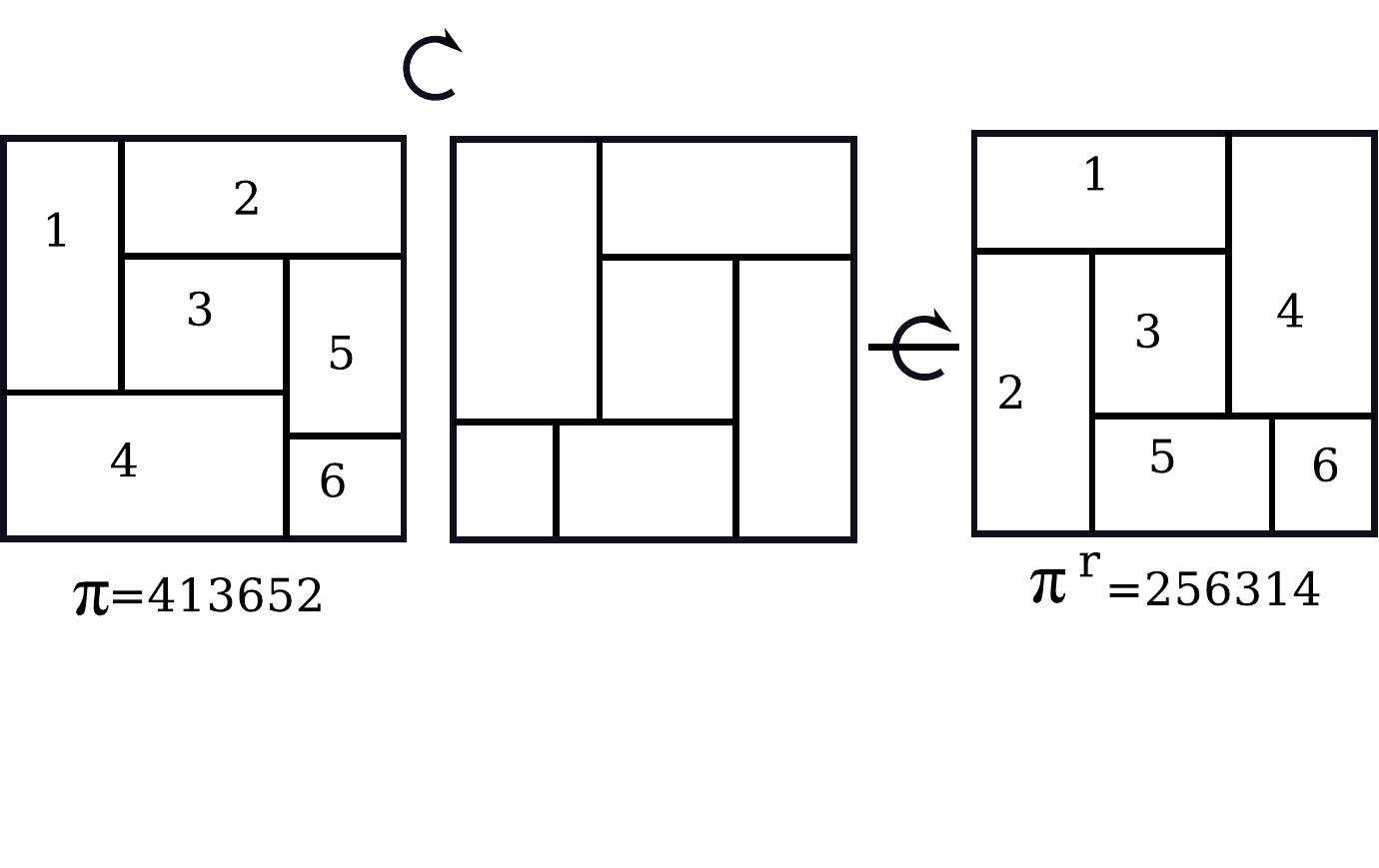}

   \caption{Obtaining a mosaic floorplan corresponding to the reverse of a Baxter permutation}
   \label{fig:geomequivrev}
   \end{figure}

\end{document}